\documentclass[11pt]{article}

\usepackage{mathtools}
\usepackage{amssymb}
\usepackage{dsfont}
\usepackage{slashed}
\usepackage{fullpage}
\usepackage[colorlinks=true,linkcolor=blue,citecolor=magenta,linktocpage=true]{hyperref}
\usepackage{titlesec}
\usepackage{tikz}
\titleformat*{\section}{\normalsize\bfseries}
\titleformat*{\subsection}{\normalsize\bfseries}
\titleformat*{\subsubsection}{\normalsize\bfseries}
\usepackage{cite}

\usepackage{hyperref}

\usepackage{subcaption}
\DeclareMathAlphabet{\bbvar}{U}{BOONDOX-ds}{m}{n}

\makeatletter
\renewcommand{\@dotsep}{10000}
\makeatother

\def\be#1\ee{\begin{align}#1\end{align}}
\def\nn{\nonumber}

\def\f{\frac}
\def\eps{\varepsilon}

\def\cD{\mathcal{D}}
\def\E{\mathcal{E}}

\def\N{\mathcal{N}}
\def\O{\mathcal{O}}

\def\cR{\mathcal{R}}

\def\T{\mathcal{T}}
\def\cT{\mathcal{T}}

\newcommand{\SL}{\mathrm{SL}}

\renewcommand{\sl}{{\mathfrak{sl}}}
\newcommand{\so}{{\mathfrak{so}}}

\renewcommand{\sl}{{\mathfrak{sl}}}

\def\rd{\textrm{d}}

\def\beq{\begin{eqnarray}}
\def\eeq{\end{eqnarray}}
\def\be{\begin{equation}}
\def\ee{\end{equation}}

\def\tA{\tilde{A}}
\def\tB{\tilde{B}}

\def\ttau{\tilde{\tau}}

\def\tLambda{\tilde{\Lambda}}

\newcommand{\R}{{\mathbb R}}

\numberwithin{equation}{section}

\begin{document}

\title{\Large{\textbf{\sffamily Symmetries and conformal bridge \\ in Schwarschild-(A)dS black hole mechanics}}}
\author{\sffamily Jibril Ben Achour\;$^{1}$, \;\; \;\; Etera R. Livine\;$^{2}$}
\date{\small{\textit{$^{1}$Arnold Sommerfeld Center for Theoretical Physics, Munich, Germany\\ $^{2}$ENS de Lyon, Univ. Claude Bernard, Lyon, France }}}

\maketitle

\begin{abstract}
We show that the Schwarzschild-(A)dS black hole mechanics possesses a hidden symmetry under the three-dimensional Poincar\'e group. This symmetry shows up after having gauge-fixed the diffeomorphism invariance in the symmetry-reduced homogeneous Einstein-$\Lambda$ model and stands as a physical symmetry of the system. It dictates the geometry both in the black hole interior and exterior regions, as well as beyond the cosmological horizon in the Schwarzschild-dS case. It follows that one can associate a set of non-trivial  conserved charges to the Schwarzschild-(A)dS black hole which act in each causally disconnected regions. In T-regions, they act on fields living on spacelike hypersurface of constant time, while in the R-region, they act on time-like hypersurface of constant radius. We find that while the expression of the charges depend explicitly on the location of the hypersurface, the charge algebra remains the same at any radius in R-regions (or time in T-regions). Finally, the analysis of the Casimirs of the charge algebra reveals a new solution-generating map. The $\sl(2,\mathbb{R})$ Casimir is shown to generate a one-parameter family of deformation of the black hole geometry labelled by the cosmological constant. This gives rise to a new conformal bridge allowing one to continuously deform the Schwarzschild-AdS geometry to the Schwarzschild and the Schwarzschild-dS solutions.  
\end{abstract}

\thispagestyle{empty}
\newpage
\setcounter{page}{1}

\hrule
\tableofcontents
%\addtocontents{toc}{\protect\setcounter{tocdepth}{2}} 
\vspace{0.7cm}
\hrule

%%%%%%%%%%%%%%%%%%%%%%%%%%%%%%%%%%%%%%%%%%%
\newpage

\section{Introduction}

Symmetries play a fundamental role in our description of physical systems. For Lagrangian field theories, Noether's theorems provide a one-to-one relation between conservation laws and symmetries of the action. At the classical level, the algebra of charges generating these symmetries constrains the structure of the solution space while at the quantum level, its irreducible representations organize the quantum theory. The conserved charges thereby identified provide a rich information on the IR sector of the theory through their algebra.

Nevertheless, identifying non-trivial charges is a subtle subject in gravity. Indeed, it is well known that any diffeomorphism stands as a pure gauge transformation associated to a trivial charge which vanishes identically in the absence of boundary. In order to associate an algebra of non-trivial charges to a gravitational field, one has to introduce a notion of boundary, being at finite distance or asymptotic. The presence of such boundary breaks the diffeomorphism gauge invariance, giving rise to non-trivial symmetries associated, via Noether's second theorem, to non-trivial surface charges \cite{Wald:1999wa, Barnich:2001jy, Barnich:2007bf, Donnelly:2016auv, Compere:2018aar, Harlow:2019yfa, Freidel:2020xyx, Odak:2021axr}. Perhaps the most well known example of this machinery in four-dimensional General Relativity (GR) is the emergence of the infinite dimensional BMS symmetry group in asymptotically flat gravitational fields \cite{Bondi:1962px, Ashtekar:1978zz, Ashtekar:1981bq, Barnich:2009se, Barnich:2011mi, Compere:2019gft, Freidel:2021fxf}. Therefore, the role of boundaries appears crucial when trying to associate an algebra of non-trivial charges to a gravitational system.

In the context of symmetry reduced GR, the case of homogeneous models  such as cosmological spacetime reveals some surprising features. For such models, the restriction to homogeneity effectively reduces the gravitational field to a mechanical system and the diffeomorphism gauge invariance of the full theory collapses to a simple time-reparametrization gauge invariance. It was noticed that, even once this gauge is fixed, the system exhibits new symmetries associated to non-trivial charges which fully encode the gravitational dynamics \cite{BenAchour:2019ufa}. In such mechanical set-up, the presence of a boundary translates into suitable cut-off scales fixing the size of the system under consideration, but its presence does not seem to play any role in the emergence of these non-trivial symmetries. Therefore, the status of these mechanical symmetries remain puzzling and begs for further exploration. The goal of this work is to discuss the realization of such hidden symmetries for the Schwarzschild-(A)dS black hole mechanics. In order to appreciate the present results, it will be useful to first review previous works on this subject.

The explicit realization of these hidden symmetries in homogenous and isotropic gravity has been investigated in a serie of works \cite{BenAchour:2019ufa, BenAchour:2020xif, Achour:2021lqq, BenAchour:2020ewm, BenAchour:2020njq}. 
As initially observed long ago in a specific context \cite{Pioline:2002qz}, the cosmological dynamics enjoys an SL$(2,\mathbb{R})$ symmetry which allows one to recast it into the mechanics of the conformal particle \cite{deAlfaro:1976vlx}. The symmetry is not a diffeomorphism but instead acts via M\"{o}bius reparametrizations of the proper time coordinate while the (truncated) gravitational field, i.e the scale factor, transforms as a primary. The conformal weight is dictated by the gauge fixing, i.e the choice of lapse, through a specific condition. While this condition is always satisfied for the flat FLRW model,  the SL$(2,\mathbb{R})$ symmetry is realized only for two possible gauge fixings when a cosmological constant or a spatial homogeneous curvature is turned on. As a consequence, the mapping onto the conformal particle initially noticed in \cite{Pioline:2002qz} can be generalized to all homogeneous and isotropic models for suitable gauge choices. It follows from these findings that one can associate to any such cosmological models a SL$(2,\mathbb{R})$ algebra of non-trivial charges which fully encodes their dynamics. See \cite{Achour:2021lqq} for details.

Another interesting outcome of these investigations is to reveal new solution-generating  mappings. Indeed, generalizing the SL$(2,\mathbb{R})$ symmetries to Diff$(S^1)$ maps, one can identify transformations associated to a constant Schwarzian cocyle which generate arbitrary constant curvature.  Such conformal transformations, dubbed conformal bridges, map the flat FLRW cosmological model to its (A)dS extensions or to the $k=\pm1$ universes  (with positive or negative homogeneously curved spatial sections) \cite{BenAchour:2020xif, Achour:2021lqq}. These transformations also appear in mechanical systems where they connect the free or the conformal particle with their Newton-Hooke extension \cite{Galajinsky:2010ry}. Such conformal bridge provides a way to dress the free system with a trapping or anti-trapping potential, connecting therefore very different physical systems which exhibit confinement or asymptotic free behavior. See \cite{Gibbons:2014zla, Inzunza:2019sct, Inzunza:2021vgt} for interesting related  investigations.

At this stage, a natural question is whether these different findings generalize to symmetry-reduced models relevant for black hole physics ? Preliminary results along that line have been obtained recently. By considering the anisotropic Kantowski-Sachs cosmological model, which describes the Schwarzschild interior mechanics, it was shown that the SL$(2,\mathbb{R})$ symmetry group found in the isotropic case is enhanced to the three dimensional Poincar\'e group SL$(2,\mathbb{R})\ltimes \mathbb{R}^3$ \cite{Geiller:2020xze}. The charge algebra associated to this symmetry group allows one to reconstruct the black hole interior dynamics. Moreover, it was shown in \cite{Geiller:2021jmg} that generalizing the symmetry group to its infinite dimensional embedding BMS$_3 =$ Diff$(S^1)\ltimes$ Vect$(S^1)$ leads to transformations generated by integrable charges, even if they are not symmetries. These results show that the structure found in isotropic cosmological models is also relevant for black hole and therefore begs  for further exploration. In the present work, we shall generalize the analysis of \cite{Geiller:2020xze} in three directions.  

First, we show that the analysis performed for the Schwarzschild black hole interior can be adapted to treat at once both the interior and the exterior regions of the black hole. To that end, we introduce a slight modification of the model considered in \cite{Geiller:2020xze} which allows us to adapt the foliation to the region of interest. While the interior dynamical region (i.e. T-region) is foliated by spacelike hypersurface of constant time, we foliate the exterior static region (i.e R-region) with timelike hypersurface of constant radius. A simple parametrization allow us to treat at once these two cases. This generalization then reveals that a symmetry group SL$(2,\mathbb{R})\ltimes \mathbb{R}^3$ also acts on each time-like hypersurface of constant radius in the exterior static region, generalizing the result of \cite{Geiller:2020xze} to the whole spacetime.

Second, we show that turning on the cosmological constant modifies the SL$(2,\mathbb{R})$ sector of the symmetry transformation while leaving the translational sector unaffected. We explicitly identify the finite symmetry transformations for the Schwarzschild-(A)dS mechanics and compute the associated conserved charges. While the charge algebra forms as expected a $\sl(2,\mathbb{R})\ltimes \mathbb{R}^3$ Lie algebra, the investigation of the Casimirs reveals some surprise. As already pointed in \cite{Geiller:2020xze}, the $\sl(2,\mathbb{R})$ Casimir coincides with the Schwarzschild mass when the cosmological constant vanishes. This is no longer true in the Schwarzschild-(A)dS case where the $\sl(2,\mathbb{R})$ Casimir is found to have a remarquable interpretation. Indeed, we find that this generator generates a one-parameter family of deformation of the black hole geometry labelled by the cosmological constant, allowing one to freely shift this key parameter. It follows that, starting from the Schwarzschild-AdS black hole solution, one can continuously deform it to the Schwarzschild solution and even switch the sign of the cosmological constant to reach the Schwarzschild-de Sitter solution. This map gives rise to a new conformal bridge between the three sectors of black hole solutions, generalizing the results found in the cosmological context in \cite{BenAchour:2020xif, Achour:2021lqq}. 

Finally, we investigate the action of the symmetry at the level of the solution space.  The existence of this Noetherian symmetry in black hole mechanics, which is an off shell statement, implies that, on shell, the physical solutions transform covariantly under finite Poincar\'e transformations. In the last section, we explicitly demonstrate that the Schwarzschild-(A)dS solution is indeed covariant under M\"{o}bius transformations and derive the induced transformations on the constants of motion involved in the solution. This demonstrates that this physical symmetry maps classical solutions of the equations of motion onto non-gauge equivalent solutions with different values of the mass and the cosmological constant.  We also present the explicit action of the conformal bridge connecting the Schwarzschild black hole geometry and its (A)dS extension. This completes the presentation of this new symmetry structure of black hole mechanics. 

The article is organized as follows. In Section~\ref{sec1}, we present the model and introduce the parametrization which allows us to treat at once both T-regions and R-regions using different foliations. To be complete, we show that our symmetry-reduced action admits indeed the Schwarzschild-(A)dS black hole family as solution. In Section~\ref{sec2}, we present the new finite symmetry transformations of the reduced action, compute the finite and infinitesimal variation of the action, and finally derive the associated Noether charges. Section~\ref{sec3} is devoted to the hamiltonian treatement of the symmetry. We present the so called extended CVH algebra\footnote{The initials C-V-H  refer respectively to the name complexifier (which corresponds for historical reasons to the integrated trace of the extrinsic curvature), the 3d volume and the hamiltonian constraint associated with a given hypersurface.}, a structure initially introduced in \cite{BenAchour:2017qpb}, and use it to rewrite the Noether charges and compute the charge algebra. The role of the $\sl(2,\mathbb{R})$ Casimir in the Schwarzschild and Schwarzschild-(A)dS case is discussed at the end of this section. Section~\ref{sec4} is devoted to the action of the symmetry on the solution space. We present the covariance of the Schwarzschild-(A)dS solution under our new symmetry group and apply explicitly the solution-generating transformation announced above. Finally, Section~\ref{sec5} is devoted to a discussion of our results and open directions.

%\begin{itemize}
%\item  Can one generalize the analysis of \cite{} for the  black hole interior to the whole spacetime ? Can one identify a symmetry group acting on the exterior region of the black hole ?  
%\item How does the symmetries transformations are affected when turning on the cosmological constant ? 
%\item Does the conformal bridge connecting the flat FLRW cosmology to its (A)dS extension reported in \cite{BenAchour:2020xif, Achour:2021lqq} can be generalized to the Schwarzschild mechanics ? In other word, can one exhibit a solution-generating map connecting the Schwarzschild black hole to its Schwarzschild-(A)dS extension ?
%\end{itemize}
 
 %irst show that the analysis of \cite{} can be easily generalized to treat both the interior and exterior regions of the black holes.

\newpage

\section{Schwarzschild-(A)dS mechanics}

\label{sec1}

In this section, we introduce the set of geometries we shall investigate throughout this work. We present the symmetry reduced action describing (A)dS mechanics and present the hidden symmetries of the gauge-fixed action. %We use this first section to set up our notation and convention. 
\subsection{The model}
Consider a spherically symmetric geometry with ADM line element 
\be
\label{ansatz}
ds^2_{\epsilon} = g_{\mu\nu} \rd x^{\mu} \rd x^{\nu} =   \epsilon \left( -N^2(x) \rd x^2 +  \gamma_{yy} (x) \rd y^2 \right) + \gamma_{\theta\theta} (x) \rd\Omega^2
\ee
where we have introduced a set of four coordinates $x^{\mu} =\{x, y, \theta, \varphi \}$ and we work with $N(x) \geqslant0$ and $\gamma_{yy}(x) \geqslant 0$ and $\epsilon=\pm1$. As we shall see, it allows us to foliate the different regions of the Schwarzschild-(A)dS geometry with spacalike or time-like slices. Indeed, it parametrizes the signature of the three dimensional hypersurface $\Sigma$ 
\be
\rd s^2_{\Sigma} = \epsilon \gamma_{yy} (x) \rd y^2  + \gamma_{\theta\theta} (x) \rd\Omega^2
\ee
 with coordinates $\{y,\theta, \varphi\}$ : for $\epsilon =+1$, $\Sigma$ is spacelike while for $\epsilon=-1$, $\Sigma$ is timelike. 
%\be
%N(x) \geqslant0\;, \qquad \gamma_{yy}(x) \geqslant 0
%\ee
Notice that the geometry is homogeneous in the sense that it depends only on the $x$ coordinate.  In what follows, we introduce the notation  $\gamma_{yy} := A^2(x)$ and $\gamma_{\theta\theta} := L^2_s B^2(x)$
% \be
% \gamma_{yy} := A^2(x) \;, \qquad \gamma_{\theta\theta} := L^2_s B^2(x)
% \ee
 where $[B] = [A] = 1$ and $L_s$ is a fiducial length scale encoding the curvature on the fiducial $2$-sphere, i.e. $L^{-2}_s$ . As usual when working with a symmetry-reduced model, we shall introduce a fiducial cell of finite size. Therefore, the range of the above coordinates is fixed to $ x \in [X_{-}, X_{+}]$, $y \in [Y_{-}, Y_{+}]$ while $\theta \in [0,\pi]$ and $\varphi \in [0,2\pi]$. In the following, it will be useful to introduce the length scale $L_0 = 4\pi (Y_{+}-Y_{-})$.

 Let us turn to the reduced action encoding the dynamics of vacuum (A)dS gravity. Using the above notation, 
% The determinant and the four dimensional Ricci scalar associated to the metric (\ref{ansatz}) are given by
%\begin{align}
%\sqrt{|g|} & = L^2_s N A B^2 \sin{\theta} \\
%\mathcal{R} & = \frac{2}{\epsilon N} \left[ \frac{\ddot{A}}{A} + \frac{2\ddot{B}}{B} + \frac{\dot{B}^2}{B^2} + \frac{2 \dot{A} \dot{B}}{AB}\right]  - \frac{2}{\epsilon N^2} \frac{\dot{N}}{N} \left[ \frac{\dot{A}}{A} + \frac{2\dot{B}}{B}\right] + \frac{2}{L^2_s B^2}
%\end{align}
the symmetry-reduced reduced action encoding the dynamics of (\ref{ansatz}) is given by
\begin{align}
S_{\epsilon} [N, A, B] & =  \int_{M} \rd^4x \sqrt{|g|} \left[ \frac{\cR - 2 \Lambda}{2 L^2_P} \right] \nn\\
\label{reduced}
& = \frac{L_0 L^2_s}{\epsilon L^2_P} \int^{X_{+}}_{X_{-}} \rd x \left[ \epsilon N A \left( \frac{1}{L^2_s} - \frac{B^2}{L^2_{\Lambda}} \right) -  \frac{A \dot{B}^2 + 2 B \dot{B} \dot{A}}{N} +  \frac{\rd}{\rd x} \left( \frac{B^2 \dot{A} + 2 A B \dot{B}}{N}\right)\right]
\end{align}
where we have introduce the cosmological length scale $L_{\Lambda} = 1/ \sqrt{\Lambda}$.
%such that $[L_0] = L$ and $\kappa = 8\pi G$ such that $[\kappa] = L^2$. We have also $[\Lambda] = L^{-2}$.
Two terms can be distinguished in the resulting action (\ref{reduced}) : the kinetic contribution in $1/N$ and the potential contribution in $N$. As expected, the parameter $\epsilon$ only affects the relative sign between these two contributions which nevertheless distinguishes the physics of the homogeneous dynamical $T$-region versus the static $R$-region. 
 %As initially noticed in \cite{BenAchour:2019ufa} and further in \cite{Geiller:2020xze}, the conformal symmetry of the reduced action is realized in the spatial slicing which has a vanishing three dimensional Ricci scalar curvature. Following \cite{Geiller:2020xze}, this slicing is achieved by a simple field redefinition of the lapse function. 
In order to investigate the symmetries of this reduced action, it will be convenient to introduce the new lapse function
\be
N\rd x = \frac{\N}{A}  \rd \eta
\ee
%The reduced action takes the form
%\begin{align}
%S_{\epsilon} [\N, A, B]& = \frac{L_0 L^2_s}{\epsilon \kappa} \int^{X_{+}}_{X_{-}}  \rd \eta \left[ \epsilon  \; \N \left( \frac{1}{L^2_s} - \frac{B^2}{L^2_{\Lambda}} \right) -  A^2 \left( \frac{\rd B}{\N \rd \eta}\right)^2 - 2 B A \frac{\rd B}{\N \rd \eta} \frac{\rd A}{\N \rd \eta}   \right. \\
%& \left. \qquad \qquad \qquad \qquad \qquad \qquad +  \frac{\rd}{\N \rd \eta} \left( AB^2 \frac{\rd A}{\N \rd \eta}+ 2 A^2 B \frac{\rd B}{\N \rd \eta} \right) \right]
%\end{align}
%such that the potential contribution coming from the 3d Ricci scalar reduces to a simple constant term $\epsilon/ L^2_s$.
such that the line element (\ref{ansatz}) takes the form
\be
\label{met}
\rd s^2 = \epsilon \left( - \frac{\N^2 \rd \eta^2}{A^2(\eta)}  + A^2(\eta) \rd y^2 \right) + L^2_s B^2(\eta) \rd \Omega^2
\ee
Implementing this field reparametrization, the reduced action reads
\begin{align}
\label{acc}
S_{\epsilon} [A, B] = \frac{L_0 L^2_s}{ \epsilon L^2_P}\int  \rd \eta \left[ \epsilon  \; \N \left( \frac{1}{L^2_s} - \frac{B^2}{L^2_{\Lambda}} \right)  - \frac{A^2 \dot{B}^2 +2 AB \dot{B} \dot{A}}{\N} \right]
\end{align}
where a dot refers now to a derivative w.r.t the coordinate $\eta$.
At this point, the transformation is just a field redefinition and no gauge fixing has been performed. In particular, the system is still gauge-invariant. Under our symmetry reduction, the diffeomorphism gauge invariance of the full theory reduces to the reparametrization
\begin{align}
\begin{aligned}
\eta \rightarrow \tilde{\eta} &= f(\eta) \\
\N \rightarrow \tilde{\N}(\tilde{\eta}) & = \N(\eta)/ \dot{f}(\eta) \\
A \rightarrow \tilde{A}(\tilde{\eta}) & = A(\eta) \\
B \rightarrow \tilde{B}(\tilde{\eta}) & = B (\eta)
\end{aligned}
\end{align}
Moreover, because of homogeneity, the boundary term will not play any role in the structure we are going to present and therefore one can drop it safely.  Nevertheless, the information on the boundary geometry is encoded in the fiducial length scales $L_0, L_s$. We shall come back in the last section on the fate of the boundary under our symmetry transformation. %Having set up our model, the action (\ref{acc}) is the final form we shall investigated in this work.  

Before going further, let us briefly discuss the explicit solution to the equations of motion. More details can be found in Appendix~\ref{appA}. Solving the dynamics for the $B$-field and $A$-field, one obtains
\begin{align}
\label{profield}
B(\tau) = \mathcal{C}_2 \left( \tau-\tau_0 \right) \;, \qquad A^2 = \epsilon \left[ \frac{(\tau - \tau_0)^2}{3L^2_{\Lambda}}  - \frac{\mathcal{C}_1}{\mathcal{C}^2_2}\frac{\tau - \tau_1}{\tau - \tau_0} \right]
\end{align}
where $(\tau_0, \tau_1)$ are constants of integration which encodes the position of the singularity and the horizons. The constant $\mathcal{C}_2$ is the velocity of the $B$-field, i.e the physical radius, and it is straigthfroward to show that $\mathcal{C}_1= L^{-2}_s$ which corresponds to the constant curvature of the 2-sphere. They are related to the scales $(L_0, L_s)$ and encode the size of the fiducial cell.  In order to make contact with the standard Schwarzschild-(A)dS solution, one can use the invariance under translation of the system and apply the transformation $\tau \rightarrow \tau + \tau_0$ and introduce the length scale 
\be
L_{M} = \tau_1 - \tau_0
\ee
Then, rescaling of the coordinates such that $\tau \rightarrow \mathcal{C}_2 \tau / \sqrt{\mathcal{C}_1} $ and $y \rightarrow \sqrt{\mathcal{C}_1} y / \mathcal{C}_2$ and the cosmological length scale as $L_{\Lambda} \rightarrow  \mathcal{C}_2 L_{\Lambda}/ \sqrt{\mathcal{C}_1} $,
%\be
%\label{rescaling}
%\tau \rightarrow \frac{\mathcal{C}_2}{\sqrt{\mathcal{C}_1}} \tau \;, \qquad y \rightarrow \frac{\sqrt{\mathcal{C}_1}}{\mathcal{C}_2} y \;, \qquad L_{\Lambda} \rightarrow  \frac{\mathcal{C}_2}{\sqrt{\mathcal{C}_1}}L_{\Lambda} 
%\ee
the solution written in proper coordinate $\rd \tau= \N \rd \eta$ becomes
\begin{align}
\label{metsol}
\rd s^2 = -  \left( 1 - \frac{L_{M}}{\tau} - \frac{\tau^2}{3L^2_{\Lambda}}\right) \rd y^2+  \left( 1 - \frac{L_{M}}{\tau} - \frac{\tau^2}{3L^2_{\Lambda}}\right)^{-1} \rd \tau^2  +  \tau^2 \rd\Omega^2
\end{align}
The resulting metric corresponds to the standard form of the Schwarzschild-(A)dS metric where $L_{M}$ is the Schwarzschild mass. The Schwarzschild-dS background corresponds to $L_{\Lambda}>0$ while its Schwarzschild-AdS counterpart is obtained by the map $L_{\Lambda} \rightarrow i L_{\Lambda}$. In the end, only the two parameters $(L_M, L_{\Lambda})$ label the solution. 

\begin{figure}[!h]
\centering
\begin{tikzpicture} [scale=2.23]

%%%%%%%%%%%%%%%%%%% PREMIER CADRAN
\draw  (0,0) -- (0,2) ;
\draw  [dashed]  (0,0) -- (2,0) ;
\draw [thick, magenta] (0,2) -- (2,0) ;
%\draw [dashed] (0,2) -- (2,2) ;
\draw [thick, magenta] (0,0) -- (2,2) ;
\draw  [dashed] (0,2) -- (2,2) ;
%%%%%%%%%%%%%%%%%%%%%%% DEUXIEME CADRAN
\draw   (2,2) -- (4,2) ;
%\draw  [dashed]  (2,0) -- (4,2) ;
\draw  [thick, magenta] (2,0) -- (4,2) ;
\draw   (2,0) -- (4,0) ;
\draw [thick, magenta] (2,2) -- (4,0) ;
\draw  (4,2) -- (4,0) ;

%%%%%%%%%%%%%%%%%%%%%%%% HYPERSURFACE
%\draw (0,0) -- (2,2) ;
\draw [brown]plot[smooth]coordinates{(2,0)(1.8,0.5)(1.7,1)(1.8,1.5)(2,2)}; %%%%%%%%%%%%% TIMELIKE
\draw [blue]plot[smooth]coordinates{(0,2)(1,1.5)(2,2)}; %%%%%%%%%%%%%%% SPACELIKE
\draw [blue]plot[smooth]coordinates{(2,2)(3,1.5)(4,2)}; %%%%%%%%%%%%%%% SPACELIKE

%%%%%%%%%%%%%%%%%%%%%%%%%%%%%%%%%%% NODES FIRST CADRAN
%\draw (2,2) node[above] {$i^{+}$};
%\draw (2,0) node[below] {$i^{-}$};
%\draw (4,2) node[above] {$i_0^{+}$};
\draw (1.9,1) node[right] {$R$};
\draw (0.95,1.7) node[right] {$T$};
%\draw (0.3,1) node[right] {$R$};
%\draw (0.3,1) node[right] {$R$};
\draw (0.95,0.35) node[right] {$T$};

%%%%%%%%%%%%%%%%%%%%%%% COORDINTES

\draw (1,2) node[above] {$\tau=0$};
\draw (3,2) node[above] {$\tau=\infty$};
\draw (1,0) node[below] {$\tau=0$};
\draw (3,0) node[below] {$\tau=\infty$};

%%%%%%%%%%%%%%%%%%%%%%%%%%%%%%% NODES SECOND CADRAN

\draw (3.7,1) node[right] {$R$};
\draw (2.95,1.7) node[right] {$T$};
\draw (0.2,1) node[right] {$R$};
\draw (2.95,0.35) node[right] {$T$};

%%%%%%%%%%%%%%%%%%%%%%%%%%% NODES HORIZONS
\draw (1.35,0.65) node[right, magenta] {$\mathcal{H}_b^{-}$};
\draw (1.5,1.45) node[below, magenta] {$\mathcal{H}_b^{+}$};

\draw (2.28,0.65) node[right, magenta] {$\mathcal{H}_c^{-}$};
\draw (2.5,1.45) node[below, magenta] {$\mathcal{H}_c^{+}$};
%\draw (0.95,0.7) node[right] {$T$};
\end{tikzpicture}
\caption{Penrose diagram of the Schwarzschild-dS black hole. The $T$-regions correspond to the region inside the black and white hole horizons $\mathcal{H}_b^{\pm}$ and the regions beyond the cosmological horizon $\mathcal{H}_c^{\pm}$. They are foliated by spacelike hypersurface of constant time (blue). The $R$-region corresponds to the region between the black hole horizon and the cosmological horizon and is foliated by timelike hypersurface (brown). }
\label{fig}
\end{figure}
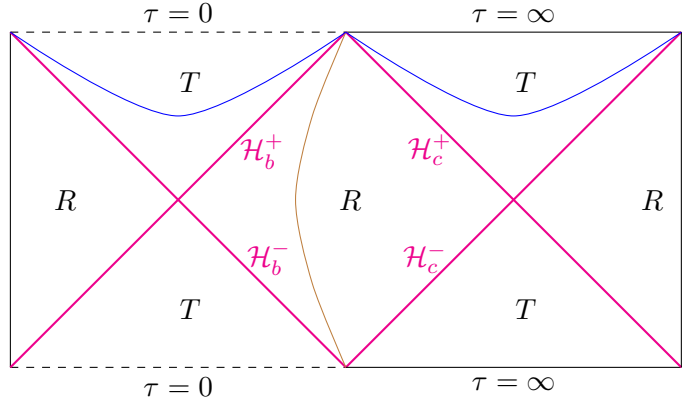

We can now comment on the role of the parameter $\epsilon = \pm 1$. Since $A^2(x) \geqslant 0$, the value of $\epsilon$ selects a range for the $\tau$-coordinate which dictates the region of spacetime covered by the metric (\ref{metsol}). In order to be concrete, let us consider the Schwarzschild-dS solution which possesses two horizons located at $\tau_{\pm}$. Its Penrose diagram is represented in Figure~\ref{fig}. In implies that
\begin{align}
 \left\{ 
    \begin{array}{ll}
        \epsilon = +1 & \mbox{then} \qquad \tau \in \;  ] \; 0, \tau_{-} ] \cup [ \tau_{+}, +\infty[  \;\;\; \text{it corresponds to a T-regions}\\
         \epsilon = - 1 & \mbox{then} \qquad \tau \in\;  [\tau_{-}, \tau_{+}]\qquad \qquad  \;\;\;\; \text{it corresponds to a R-region}
    \end{array}
\right.
\end{align}
Therefore, the parameter $\epsilon$ allows us to treat in a simple way the different causally disconnected regions of the Schwarzschild-dS geometry (and its AdS counter-part).% The Penrose diagrams of the Schwarzschild-dS backgrounds and their different $T$-regions and $R$-regions are represented in Figure~\ref{fig} . 
Having presented in detail our set-up, we are now ready to explore the hidden symmetries of the Schwarzschild-(A)dS mechanics.

\newpage

\subsection{Symmetries of the gauge-fixed action}
\label{sec2}

In this section, we present the hidden symmetries of Schwarzschild-(A)dS mechanics. We introduce for convenience the dimensionless quantity $\kappa = L_0L^2_s / L^3_P$ which encodes the ratio between the effective size of the system, i.e the IR cut-off, and the Planck volume, i.e the UV cut-off our our system. 

Now, consider the symmetry reduced action (\ref{acc}) and let us fix the gauge by reabsorbing the lapse field through the $\tau$-coordinate defined by $\rd \tau = \N \rd \eta$.
%\be
%\label{tau}
%\rd \tau = \N \rd \eta
%\ee
The gauge-fixed action reads
\begin{align}
\label{gfac}
S_{\epsilon} [A, B] = \frac{\kappa}{ \epsilon}\int  \rd \tau \left[ \epsilon  \left( \frac{1}{L^2_s} - \frac{B^2}{L^2_{\Lambda}} \right)  - A^2 \dot{B}^2 - 2 AB \dot{B} \dot{A} \right]
\end{align}
%{\Ji At this stage, the mechanical system we obtain is no longer gauge-invariant under $\tau$-reparametrization and the hamiltonian of the system is a priori non vanishing. In order to reproduces the black hole dynamics, and recover the equation of motion coming from the variation of the lapse, one has to restrict the above system to the case $H=0$.} 
%
Let us make two important remarks at this stage.
First, we started from an action principle $S[\eta,\N(\eta),A(\eta),B(\eta)]$, gauge invariant under time reparametrizations, to a gauge-fixed action $S[\tau,A(\tau),B(\tau)]$ where the lapse $\N$ has been re-absorbed in the $\tau$-coordinate. The equation of motion of the original gauge-invariant action with respect to the variation of the lapse $\N$ imply that its Hamiltonian is of the form $H_{\eta}[\N,A,B]=\N{\cal H}[A,B]$ and therefore vanishes on-shell, i.e. that we have a Hamiltonian constraint ${\cal H}=0$.  If one starts directly from the gauge-fixed action $S[\tau,A(\tau),B(\tau)]$, without remembering where it comes from, one computes a Hamiltonian $H_{\tau}[A,B]={\cal H}[A,B]$, which does not necessarily vanishes. One should nevertheless remember that there is an implicit dependence on the lapse field $\N$ hidden in the $\tau$-coordinate, through its definition $\rd \tau = \N \rd \eta$, and that this leads to a non-trivial equation of motion with respect to lapse variations $\delta\N$. In simpler words, the classical solutions for the gauge-fixed action $S[\tau,A(\tau),B(\tau)]$ are classical solutions for the gauge-invariant action principle $S[\eta,\N(\eta),A(\eta),B(\eta)]$ when and only when the Hamiltonian vanishes $H_{\tau}[A,B]={\cal H}[A,B]=0$.

The second point concerns the status of the constant term $\eps L_{s}^{-2}$ in the gauge-fixed action $S[\tau,A(\tau),B(\tau)]$. Constant terms are irrelevant for the equations of motion, so we could simply remove it from the Lagrangian. However, one should keep in mind that it is actually a term $\eps L_{s}^{-2}\,\rd\tau$, with some implicit dependence on the lapse. Indeed it actually comes from a term $\N \eps L_{s}^{-2}\,\rd\eta$ in the original gauge-invariant action. Such a term is clearly not constant and affects the equation of motion derived from lapse variations. In fact, it shifts the Hamiltonian constraint, from $H_{\eta}[\N,A,B]=\N{\cal H}[A,B]$ to $H_{\eta}[\N,A,B]=\N\,({\cal H}[A,B]-\eps L_{s})$. Thus, the role of the apparently  constant term $\eps L_{s}^{-2}$ is essential: it changes the value of the Hamiltonian $H_{\tau}$ of the gauge-fixed system.

We are now ready to discuss the hidden symmetries of the gauge-fixed action (\ref{gfac}). Consider the following finite transformations
\begin{align}
\begin{aligned}
\label{sym1}
\tau \rightarrow \tilde{\tau} & = f(\tau) \\
%\label{sym2}
B(\tau) \rightarrow \tilde{B}(\tilde{\tau}) & = \dot{f}^{1/2} B(\tau) \\
%\label{sym3}
A^2(\tau) \rightarrow \tilde{A}^2(\tilde{\tau}) & = A^2(\tau) + \chi(\tau) 
\end{aligned}
\end{align}
where the functions $\left(f(\tau), \chi(\tau), h(\tau) \right)$ are given by 
\begin{align}
\label{f}
f(\tau) & = \frac{a \tau + b}{c\tau + d} \;,  \\
\label{corr}
\chi (\tau)&:= 2 h \frac{\dot{B}}{B} - \dot{h} - \frac{2 h\circ f}{\dot{f}} \left[ \frac{\dot{B}}{B} + \frac{\ddot{f}}{2\dot{f}}\right] + \dot{h} \circ f \;, \\
\label{h}
h (\tau) & :=\frac{\epsilon}{3L^2_{\Lambda}} \tau^3  \;,
%h_{f} (\tau) & := h \circ f (\tau) = \frac{\epsilon}{3L^2_{\Lambda}} f(\tau)^3
\end{align}
The constants $(a,b,c,d)$ are real and satisfy $ad-bc \neq 0$. Notice that these transformations preserve the gauge-fixing. A lengthy computation detailed in Appendix~\ref{appB} reveals that the finite transformation of the action reads
\begin{align}
\label{varacc}
\Delta S_{\epsilon}  
& = \epsilon \kappa L_P \int \rd \tau  B^2 \left\{ \frac{1}{2} \text{Sch}[f]  \left [  A^2 + \chi - 4 \dot{h}\circ f \right]  + \frac{h\circ f}{\dot{f}}  \; \frac{\rd}{\rd \tau} \text{Sch}[f] \right\}  + \epsilon \kappa L_P \int \rd \tau \frac{\rd F}{\rd \tau} 
%& +  \epsilon c L_P \int \rd \tau B^2 \left\{  \frac{\epsilon}{L^2_{\Lambda}}  \left[ \left( \frac{\dddot{h}}{2} - 1\right) -  \dot{f}^2 \left(  \frac{\dddot{h}\circ f}{2} - 1\right) \right] \right\} \\
%& +  \epsilon c L_P \int \rd \tau \frac{\rd}{\rd \tau} \left\{ \frac{\epsilon}{L^2_s} ( f - \tau) + \left[  \frac{\rd^2}{\rd \tau^2} \left( h - \frac{h\circ f}{\dot{f}}\right) -\frac{\ddot{f}}{\dot{f}} (A^2 + \chi )   \right] \frac{B^2}{2} -   \left( h - \frac{h\circ f}{\dot{f}}\right) \dot{B}^2   \right\}
\end{align}
where the expression of $F$ is given by (\ref{F}). For a M\"{o}bius reparametrization, the Schwarzian defined by (\ref{sch}) vanishes and the variation reduces to a total derivative term, showing that it is indeed a Noether symmetry of the gauge-fixed action. The coefficient in front of the anomalous term, which contains the Schwarzian and its derivative, is identified as the central charge. It coincides with our dimensionless constant $\kappa$ fixing the size of the system.

Let us now discuss the Noether charges generating this symmetry. Considering an infinitesimal version of this transformation, i.e. $f(\tau) = \tau + \xi(\tau)$, the M\"{o}bius transformation can be decomposed into
\begin{align}
\label{ep}
\xi(\tau) = \left\{
    \begin{array}{ll}
        \sigma  & \mbox{for translation : $a = d =1$, $c=0$ and $b = \sigma$ }  \\
        \sigma \tau & \mbox{for dilatation :  $b = c =0$, $c= 1$ and $a = \sigma$} \\
        \sigma \tau^2  & \mbox{for special conformal transformation  :  $a=d=1$, $b=0$ and $c = \sigma$}
    \end{array}
\right.
\end{align}
The Noether charges generating these infinitesimal transformations are respectively given by
\begin{align}
\label{ch1}
Q^{\epsilon}_{+} & = -  \epsilon \kappa L_{P} \left\{  A^2 \dot{B}^2 + 2 A \dot{A} B \dot{B} \right\} + \frac{\kappa L_P}{L^2_{\Lambda}} \left\{ 2\tau B \dot{B} - \tau^2 \dot{B}^2\right\}\\
\label{ch2}
Q^{\epsilon}_{0} \; & = \epsilon \kappa L_{P} \left\{  A \dot{A} B^2 + B \dot{B} A^2 - \tau  \left( A^2 \dot{B}^2 + 2 A \dot{A} B \dot{B}\right)\right\}  + \frac{\kappa L_P}{L^2_{\Lambda}} \left\{ 2 \tau^2 B \dot{B} - \tau B^2 - \frac{2}{3} \tau^3 \dot{B}^2 \right\}\\
Q^{\epsilon}_{-} & = \epsilon \kappa L_{P} \left\{2\tau \left( A \dot{A} B^2 + B \dot{B} A^2\right) -  A^2B^2 - \tau^2 \left( A^2 \dot{B}^2 + 2 A \dot{A} B \dot{B}\right)  \right\} \nn \\
\label{ch3}
& \qquad \qquad \qquad \qquad \qquad \qquad \qquad \qquad \qquad \qquad \qquad + \frac{\kappa L_{P}}{L^2_{\Lambda}} \left\{  \frac{4}{3} \tau^3 B \dot{B} - \tau^2 B^2 - \frac{\tau^4}{3} \dot{B}^2\right\}
\end{align}
The conservation of these charges is easily checked. See (\ref{consq}) in Appendix~\ref{appB}. The expression of these charges reduces to the results of \cite{Geiller:2020xze} for $L_{\Lambda} \rightarrow +\infty$ and $\epsilon = +1$. They generate the SL$(2,\mathbb{R})$ sector of the symmetry of the Schwarzschild-(A)dS black hole mechanics.
%%%%%%%%%%%

Let us now discuss the second symmetry of the gauge-fixed action. Consider the finite transformations
\begin{align}
\begin{aligned}
\label{sym11}
 \tau \rightarrow \tilde{\tau}  & = \tau \\
%\label{sym12}
A^2(\tau) \rightarrow  %\tilde{A}(\tilde{\tau}) \;, \qquad \text{such that } \qquad
 \tilde{A}^2(\tilde{\tau})  & =  A^2 (\tau)  + 2h \frac{ \dot{B} }{B} - \dot{h} \\
%\label{sym13}
B \rightarrow \tilde{B}(\tilde{\tau})  & =  B (\tau)
\end{aligned}
\end{align}
%and let us choose $h(\tau)$ of the form
%\be
%\label{hh}
%h(\tau) := \delta +\gamma \tau+ \beta \tau^{2}
%\ee
A direct computation shows that
\begin{align}
\label{varT}
\Delta S % = \frac{L_0L^2_s}{\epsilon L^2_P} \int \rd \tau \left[ \frac{\epsilon}{L^2_s} - \frac{\epsilon}{L^2_{\Lambda}} B^2 - \dot{B}^2 \left( A^2 + 2 h \frac{\dot{B}}{B} - \dot{h} \right) - B \dot{B} \left( 2 A \dot{A} + 2 \dot{h} \frac{\dot{B}}{B} + 2h \frac{\ddot{B}}{B} - 2 h \frac{\dot{B}^2}{B^2} - \ddot{h}\right)\right] \\
%& = S + \frac{L_0L^2_s}{\epsilon L^2_P} \int \rd \tau \left[ - \dot{h} \dot{B}^2 - 2 h \dot{B} \ddot{B} + \ddot{h} B \dot{B}\right] \\
& =  \epsilon \kappa L_{P} \int \rd \tau \left[ \frac{\rd}{\rd \tau} \left( \ddot{h} B^2 - h \dot{B}^2 \right) - \dddot{h} \frac{B^2}{2} \right]
\end{align}
Therefore, for $h(\tau) := \delta +\gamma \tau+ \beta \tau^{2}$ where $(\delta, \gamma, \beta)$ are real constants, this variation reduces to a total derivative and one obtains a second symmetry of the action. Moreover, choosing instead the function $h(\tau)$ such that $\dddot{h}$ is a constant, the above transformation is not a symmetry but instead generates a new term in the gauge-fixed action which shifts the value of the cosmological constant via the second term in (\ref{varT}). This reveals the existence of a solution-generating map. We shall discuss this point in detail at the end of Section~\ref{cazou}. Coming back the the symmetry transformation with $\dddot{h}=0$, the Noether charges generating the infinitesimal version of this symmetry are easily computed and read
\begin{align}
\label{cht1}
T_{+}^{\epsilon} & =  \epsilon \kappa L_{P} \dot{B}^2 \\
T_{0}^{\epsilon} & =  \epsilon \kappa L_{P} \left\{ \tau \dot{B}^2 - B \dot{B} \right\}  \\
\label{cht3}
T_{-}^{\epsilon} & =  \epsilon \kappa L_{P} \left\{ \tau^2 \dot{B}^2 - 2\tau B \dot{B} + B^2\right\} 
\end{align}
They are not affected by the presence of the cosmological constant and therefore coincide with the ones found in \cite{Geiller:2020xze} for the Schwarzschild black hole. However, with our new parametrization, the above expressions for the translational charges corresponds both to the interior and the exterior regions of the black hole. The conservation of these charges is easily obtained. See (\ref{consT}) in Appendix~{\ref{appB}}. This second set of charges generate the translation sector of the symmetry group. It follows from this section that the Schwarzschild-(A)dS black mechanics possesses a set of hidden symmetries under the three dimensional Poincar\'e group SL$(2,\mathbb{R})\times \mathbb{R}^3$.

\section{Poincar\'e structure of Schwarzschild-(A)dS mechanics}
\label{sec3}
%%%%%%%%%%%

In this section, we show that the symmetry identified above translates at the phase space level in the so called extended CVH algebra. We further show that the Noether charges computed in Section~\ref{sec2} form as expected an $\sl(2,\mathbb{R})\ltimes \mathbb{R}^3$ algebra which upgrades the extended CVH algebra to a $\tau$-dependent structure. We conclude this section by the analysis of the Casimirs of the system and identify the  $\sl(2,\mathbb{R})$ Casimir as generating a one-parameter family of deformation of the dynamics labelled by the cosmological constant.

\subsection{Hamiltonian formulation and physical observables}

Consider therefore the reduced action (\ref{acc}).
%\begin{align}
%S_{\epsilon} [A, B] = \epsilon \kappa L_P\int  \rd \eta \left[ \epsilon \; \N \left( \frac{1}{L^2_s}  -  \frac{B^2}{L^2_{\Lambda}} \right)  - \frac{A^2 \dot{B}^2 +2 AB \dot{B} \dot{A} }{\N}\right]
%\end{align}
The conjugate momentum are given by
\begin{align}
P_A  = - 2 \epsilon \kappa L_P  \frac{A B \dot{B} }{\N} \;, \qquad P_B = - 2 \epsilon \kappa L_P  \frac{A}{\N} \left( A \dot{B} + B \dot{A}\right) \;, 
\end{align}
such that $[P_A] =[P_B] = [A]= [B] =1$ and $\{ P_A, A\} = \{ P_B, B \} =1$.
%\be
%\{ P_A, A\} = \{ P_B, B \} =1
%\ee
%At this stage, we have four length scales involved, namely $(L_0, L_s, L_{\Lambda}, L_{P})$. To simplify our notation, it is convenient to introduce the new length scales $(\tilde{L}_0, \tilde{L}_s,\tilde{L}_{\Lambda})$ given by
%\be
%\label{scale}
%\tilde{L}_0 = \frac{L_0 L^2_{s}}{L^2_P} \;, \qquad \tilde{L}_s = \frac{L^2_s}{\tilde{L}_0} = \frac{L^2_P}{L_0} 
%\;, \qquad \tilde{L}_{\Lambda} = \frac{L^2_{\Lambda}}{\tilde{L}_0}
%\ee 
%which will enter explicitly in the hamiltonian of the system.
Then the reduced action takes the form
\be
S_{\epsilon} = \int \rd \eta \left[ P_A \dot{A} + P_B \dot{B} - \N H\right]
\ee
where the physical hamiltonian reads
\begin{align}
\label{ham0}
H & =  \frac{\epsilon }{2 \kappa L_P} \left[ \frac{P^2_A}{2B^2} - \frac{P_A P_B}{AB} \right] + \kappa L_{P} \left[   \frac{B^2}{ L^2_{\Lambda}} - \frac{1}{L^2_s} \right] 
\end{align}
such that $[H] = L^{-1}$. The first term contains the kinetic contribution, the second term contains the contribution of the curvature induced by the cosmological constant while the last term encodes the contribution of the curvature induced by the 2-sphere.
Let us now introduce a new set of canonical variables given by
\begin{align}
V_1 &= B^2\;, \;\;\;\; \qquad P_1 =  \frac{1}{2B} \left(  P_B - \frac{A P_A}{B} \right)\;,  \\
V_2 &= \frac{A^2 B^2}{2}, \qquad P_2 = \frac{ P_A}{AB^2}
\end{align}
We have that $[P_1]=[P_2]=[V_1] =[V_2] =1$ and $\{ V_1, P_1\} = \{ V_2, P_2 \}  = 1$ while all other brackets vanish. With this new set of canonical variables, the hamiltonian takes the form
\begin{align}
\label{ham}
H =   - \frac{ \epsilon }{\kappa L_P} \left[  V_1 P_1 P_2 + \frac{1}{2} V_2 P^2_2 \right]  +  \kappa L_{P} \left[ \frac{V_1}{L^2_{\Lambda}} - \frac{1}{L^2_s}\right]
\end{align}
which corresponds to the notation introduced in \cite{Geiller:2020xze}.
Let us now discuss the physical observables of the system. A straightforward computation shows that the Dirac observables of the vacuum (A)dS system are given by the two phase space functions
\begin{align}
\label{obs1}
\O_1 & := V_1 P_1 -  \frac{2}{3} \frac{\kappa^2 L^2_P}{ L^2_{\Lambda}} \epsilon \frac{V_1}{P_2} \;, \\
\label{obs2}
  \O_2 & :=  \frac{V_1 P^2_2}{2}
\end{align}
which satisfy $\{ \O_1, H\} = 0 = \{ \O_2, H\}$. Notice that $[\O_2] = [\O_1] =1$. Computing these observables for the Schwarzschild-(A)dS solution using the metric functions (\ref{profield}) with the associated value for $\epsilon$, one can show that
\begin{align}
\O_1= \frac{1}{2} \kappa L_P \mathcal{C}_1 (\tau_1 - \tau_0) =  \frac{L_0 L_{M}}{2 L^2_P}  \;, \qquad 
\O_2  = 2 \kappa^2 L^2_P \mathcal{C}^2_2 = \frac{2L^2_0 L^2_s}{L^2_P} \mathcal{C}^2_2
\end{align}
Therefore,  $\O_1$ encodes the mass of the system given by the length scale $L_M= \tau_1 - \tau_0$. On the other hand, $\O_2$ encodes the effective size of the system. One can thus fix $\mathcal{C}_2 = L^{-1}_P$ without loss of generality such that $\O_2$ coincides with the squared central charge $\kappa$. Now, at the phase space level, the symmetry of the system becomes manifest in the existence of the so called CVH algebra which we now present.

\subsection{The extended CVH algebra }

Consider now the shifted hamiltonian given by
\be
\tilde{H} = H + \frac{\kappa L_{P}}{L^2_s} = -  \frac{\epsilon}{\kappa L_P} \left[  V_1 P_1 P_2 + \frac{1}{2} V_2 P^2_2 \right]+  \frac{\kappa L_{P}}{L^2_{\Lambda}} V_1
\ee
and let us introduce the phase space function $C$ defined by
\be
C := \kappa L_{P} \{ V_2, \tilde{H}\} = -  \epsilon \left( V_1 P_1 +  V_2 P_2\right)
\ee
With the phase space variables $(V_1, V_2)$, they satisfy the following brackets
\begin{align}
\begin{aligned}
\{ C, V_2\} & = + \epsilon V_2\;, \qquad  \{C, \tilde{H}\}  =- \epsilon \tilde{H} + \epsilon  \frac{2\kappa L_P}{L^2_{\Lambda}} V_1\;, \qquad \{ V_2, \tilde{H}\}  = \frac{C}{\kappa L_{P}}
\end{aligned}
\end{align}
where the parameter $\epsilon = \pm 1$ enters explicitly. We see that when $L_{\Lambda} \rightarrow +\infty$, the algebra closes which corresponds to the results of \cite{Geiller:2020xze}. This algebra is known as the CVH algebra and was initially found and investigated in cosmological models \cite{BenAchour:2019ufa, BenAchour:2020xif, Achour:2021lqq}. Here, we see that  this CVH algebra is no longer closed for the Schwarzschild-(A)dS system as it inherits a term linear in $V_1$ which stands out of the triplet $(C, V_2, \tilde{H})$. Therefore, one has to look for an extended version of it. To that end, let us define the quantity
\be
D : = \kappa L_{P} \{ V_1 , \tilde{H} \} = - \epsilon V_1 P_2
\ee
Together with the other elements, it satisfies the following commutation relations
\begin{align}
\begin{aligned}
\{ D, V_1\} = \{ D, C\} & = 0\;,  \qquad \{ D, V_2 \}  = + \epsilon V_1 \;, \;\;\;\; \{ C, V_1 \}  = + \epsilon V_1 \;,\;\;\;\;  \{ D, \tilde{H}\}  =  \frac{\O_2}{\kappa L_{P}}\;, 
\end{aligned}
\end{align}
where $\O_2= V_1 P^2_2/2$ is the Dirac observable introduced previously in (\ref{obs2}). The remaining brackets are given by
\begin{align}
\{  \O_2, C \} = + \epsilon \O_2 \;,\qquad \{ \O_2, V_2\} =  \epsilon D \;, \qquad \{ O_2 , V_1 \} = \{ \O_2, D\} =0
\end{align}
which allows one to close the algebra. This structure is therefore generated by six generators $( C, V_2, \tilde{H}, V_1, D, \O_2 )$. Notice that except $\tilde{H}$, none of the generators depend on the cosmological constant $L_{\Lambda}$. Therefore, their expressions are the same for the Schwarzschild case. 

\subsection{Algebra of Noether charges}

\label{sec43}

Using these phase space functions, one can rewrite the $\text{sl}(2,\mathbb{R})$  Noether charges derived in Section~\ref{sec2} as
\begin{align}
\label{c1}
Q^{\epsilon}_{+} & = \left(  \tilde{H} - \frac{\kappa L_P}{L^2_{\Lambda}} V_1 \right) + \frac{1}{L^2_{\Lambda}} \left( \tau D - \tau^2 \frac{\O_2}{2\kappa L_P}\right) \\
Q^{\epsilon}_{0} \; & =  \tau \left(  \tilde{H} - \frac{\kappa L_P}{L^2_{\Lambda}} V_1 \right) + \epsilon C + \frac{1}{L^2_{\Lambda}} \left( \tau^2 D - \tau \kappa L_P V_1 - \frac{1}{3} \tau^3  \frac{\O_2}{\kappa L_P} \right) \\
 Q^{\epsilon}_{-}  & =  \tau^2 \left(  \tilde{H} - \frac{\kappa L_P}{L^2_{\Lambda}} V_1 \right) + 2\epsilon \tau C - 2 \epsilon \kappa L_P V_2 + \frac{1}{L^2_{\Lambda}} \left( \frac{2}{3} \tau^3 D - \tau^2 \kappa L_P V_1 - \frac{\tau^4}{6} \frac{\O_2}{\kappa L_P}\right)
\end{align}
while the translation Noether charges can be recast as
\begin{align}
T_{+}^{\epsilon}  =  \frac{\epsilon}{2\kappa L_P} \O_2 \;, \qquad T_{0}^{\epsilon}  =  \frac{\epsilon}{2}  \left\{ \tau  \frac{\O_2}{\kappa L_P} - D\right\}  \;, \qquad 
T_{-}^{\epsilon}  =  \epsilon  \left\{ \tau^2 \frac{\O_2}{2\kappa L_P} - \tau D + \kappa L_P V_1\right\} 
\end{align} 
The conservation and the algebra satisfied by these charges are easily checked. %which reads 
%\be
%\frac{\rd Q^{\epsilon}_{-}}{\rd \tau} = \{ Q^{\epsilon}_{-}, \tilde{H}\} + \frac{\partial Q^{\epsilon}_{-}}{\partial \tau} = 0
%\ee
%for the charge $Q^{\epsilon}_{-}$ while the same conservation equation holds for the other charges. 
The $\text{sl}(2,\mathbb{R})$ sector is given by
\be
 \{ Q^{\epsilon}_{+}, Q^{\epsilon}_{-}\} = 2 Q^{\epsilon}_0\;, \qquad \{ Q^{\epsilon}_0, Q^{\epsilon}_{-}\} = Q^{\epsilon}_{-} \;, \qquad \{ Q^{\epsilon}_0, Q^{\epsilon}_{+}\} = - Q^{\epsilon}_{+}
\ee
while the charges generating the translations commute
%\begin{align}
%\{  T_{+}^{\epsilon}, T_{-}^{\epsilon}\} = \{ T_{0}^{\epsilon}, T_{+}^{\epsilon} \}= \{ T_{0}^{\epsilon}, T_{-}^{\epsilon}\} =0 \;, 
%\end{align}
as expected.  The cross brackets are given by
\begin{align}
\begin{aligned}
& \{ Q^{\epsilon}_{-}, T_{+}^{\epsilon} \} = 2 T_{0}^{\epsilon} \;, \qquad \{Q^{\epsilon}_{+}, T_{+}^{\epsilon} \} =  0 \;, \qquad \;\;\; \{ Q^{\epsilon}_{0}, T_{+}^{\epsilon}\} = - T_{+}^{\epsilon} \;, \\
& \{ Q^{\epsilon}_{-}, T_{-}^{\epsilon} \} = 0 \;, \qquad  \;\;\;\; \{ Q^{\epsilon}_{+} ,  T_{-}^{\epsilon}\} = 2 T^{\epsilon}_0 \;, \qquad \{  Q^{\epsilon}_{0} ,  T_{-}^{\epsilon}\} =  T_{-}^{\epsilon} \;, \\
&  \{ Q^{\epsilon}_{-}, T_{0}^{\epsilon} \} = - T^{\epsilon}_{-} \;, \qquad  \{ Q^{\epsilon}_{+} ,  T_{-}^{\epsilon}\} = T^{\epsilon}_{+} \;, \qquad \{  Q^{\epsilon}_{0} ,  T_{0}^{\epsilon}\} = 0\;, 
\end{aligned}
\end{align}
It is interesting to note that while the expression of the conserved charges depend explicitly on the parameter $\epsilon$ as well as on the $\tau$-coordinate, the brackets of the charge algebra do not. Therefore, the charge algebra is the same in both $T$-region ($\epsilon=+1$) and $R$-region $(\epsilon=-1)$ and keep the same form on any hypersurface and encodes the geometry at any radial position, close to the horizon, deep inside the trapped region, or far away from the black hole horizon. In the dS case, it also holds beyond the cosmological horizon. %Therefore, the structure allows one to associate a non-trivial charge algebra to the Schwarzschild-(A)dS black hole, being asymptotically de Sitter or anti-de Sitter. 

\subsection{$2+1$ Poincar\'e algebra}

Now, we would like to show how the above extended CVH algebra can be organized into a $2+1$ Poincar\'e algebra. Let us first focus on the $\sl(2,\mathbb{R})$ sector. The phase space functions $(C, V_2, V_1, \tilde{H})$ can be combined to form the following boosts and rotation generators
\begin{align}
\begin{aligned}
K_y & = C \;, \\
 K_x & = \sqrt{\frac{\kappa L_{P}}{2}} \left[ \frac{\epsilon V_2}{\sqrt{\sigma  L_P}}  +  \sqrt{\sigma L_{P}} \left( \tilde{H} -  \frac{ \kappa L_{P}}{L^2_{\Lambda}} V_1 \right)\right] \;, \\
 J_z &=   \sqrt{\frac{\kappa L_{P}}{2}} \left[ \frac{ V_2}{\sqrt{\sigma L_P}}  -  \epsilon \sqrt{\sigma L_P} \left( \tilde{H}  -  \frac{ \kappa L_{P}}{L^2_{\Lambda}} V_1 \right) \right] 
   \end{aligned}
  \end{align}
%  The identification to the standard boosts and rotation generators forming $\sl(2,\mathbb{R})$ algebra depends on the value of the parameter $\epsilon$ and therefore if one considers the exterior or interior region. It is given by 
%    \begin{align}
%  F_0 = K_y\;, \qquad F_{1} = \left\{
%    \begin{array}{ll}
%        K_x & \mbox{if } \;\;\; \epsilon = +1 \\
%        J_z & \mbox{if} \;\;\; \epsilon = -1
 %   \end{array}
%\right.
% \qquad    F_{2} = \left\{
%    \begin{array}{ll}
%        J_z & \mbox{if } \;\;\; \epsilon = +1 \\
%        K_x & \mbox{if} \;\;\; \epsilon = -1
%    \end{array}
%    \right.
%  \end{align}
where the parameter $\sigma \in \mathbb{R}$ is free.
Notice that the generators $K_x$ and $J_z$ both depend on the parameter $\epsilon$. The triplet $(K_x, K_y, J_z)$ statisfy the standard  $\sl(2,\mathbb{R})$ commutation relations given by
\begin{align}
& \{ J_z, K_x\}  = K_y \;, \qquad  \{ J_z, K_y\} =  -  K_x \;, \qquad \{ K_x, K_y \}  = - J_z
\end{align}
The $\sl(2,\mathbb{R})$ Casimir generator is given by
\begin{align}
\begin{aligned}
\label{caz}
\T_0  & =  J^2_z - K^2_x - K^2_y \\
& = 2 \epsilon  \kappa L_{P} V_2 \left( \tilde{H} - \frac{\kappa L_P}{L^2_{\Lambda}} V_1 \right) - C^2 =  - V_1^2 P^2_1 <0
\end{aligned}
\end{align}
Therefore, the $\sl(2,\mathbb{R})$ Casimir is independent of $\epsilon$ and always negative. From the quantization point of view, it selects the continuous serie of the irreducible representation of $\SL(2,\mathbb{R})$. When $L_{\Lambda} \rightarrow + \infty$, this Casimir coincides with the observable $\O^2_1$ and thus with the squared Schwarzschild mass, which implies that the mass spectrum is continous. When $L_{\Lambda}$ is finite, the Casimir is no longer a Dirac observable. Its role will be discussed in the next section.

Now let us focus on the translation sector. The phase space functions $(D, V_1, \O_2)$ can be organized as
  \begin{align}
    \begin{aligned}
P_x & = - D \;, \\
 P_y & =  \epsilon \sqrt{\frac{\kappa L_{P}}{2}} \left[ \frac{V_1}{\sqrt{\sigma L_P}} -  \sqrt{\sigma L_P} \frac{\O_2}{\kappa L_p} \right]  \;, \\
  P_z & =   \sqrt{\frac{\kappa L_{P}}{2}} \left[ \frac{ V_1}{\sqrt{\sigma L_P}} + \sqrt{\sigma L_P} \frac{\O_2}{\kappa L_p} \right] \;, 
    \end{aligned}
\end{align}
where the parameter $\sigma \in \mathbb{R}$ is free and $\O_2$ is the Dirac observable introduced in (\ref{obs2}). Notice that $P_y$ depends on the parameter $\epsilon$.
It is straightforward to check that the generators $(P_x, P_y, P_z)$ commute as expected
%\begin{align}
%& \{ P_x, P_y\}  =  0 \;, \qquad \qquad \{ P_x, P_z\} =  0 \;, \qquad \qquad  \{ P_y, P_z \}  =  0
%\end{align}
and that the Casimir operator associated to this sector vanishes 
\be
\mathcal{T}_1  =  P^2_z - P^2_x - P^2_y =0 
\ee
It describes the translation sector. We can now look at the brackets between the $\sl(2,\mathbb{R})$ generators and the translation generators. Step by step, the remaining brackets are given by
\begin{align}
\begin{aligned}
& \{ K_x, P_x\} = P_z \;, \qquad \{ K_x, P_y\}  =0 \qquad \;\; \; \;\; \{ K_x, P_z\}  =  P_x\; \\
&  \{ K_y, P_x\}  =  0 \;, \qquad \;\; \{ K_y, P_y \} =  P_z \qquad \;\;\; \; \{ K_y, P_z\} =  P_y\;, \\
&  \{J_z, P_x \} =  P_y   \;, \qquad\;\;   \{ J_z, P_y\} = -  P_x\;, \qquad   \{ J_z, P_z\} = 0\;,
\end{aligned}
\end{align}
Before closing this section, let us finally compute the last Casimir of this Poincar\'e algebra. It is given by
\begin{align}
\mathcal{T}_2  = J_z P_z  + K_y P_x - K_x P_y = 0
\end{align}
Again, notice that the brackets of this 2+1 Poincare algebra do not depend on the parameter $\epsilon$.
This concludes the presentation of the algebraic structure of the Schwarzschild-(A)dS mechanics. It follows that the Schwarzschild-(A)dS black hole can be characterized by three invariant numbers which label its classical states and correspond to the allowed values of the three Casimirs $(\T_0, \mathcal{T}_1, \mathcal{T}_2)$, the only non-vanishing one being $\T_0$.

%An interesting difference from \cite{Geiller:2020xze} is the role played by the Casimir of the $\textbf{sl}(2)$ sub-algebra which is not more a Dirac observable of the system. In the next section, we clarify its role and show how it allows one to generate a cosmological constant.

\subsection{Role of the $\sl(2,\mathbb{R})$ Casimir}

\label{cazou}

In this section, we discuss the role of the non-vanishing Casimir $\T_0$ for the Schwarzschild and Schwarzschild-(A)dS black holes. %for the different value of the cosmological constant $\Lambda$. For $\Lambda=0$, we show that the Casimir is a Dirac observable which admits a discrete (resp. continuous) spectrum in the the exterior (resp. interior) regions. For $\Lambda\neq 0$, the Casimir is no longer a Dirac observable but generates the flow along the cosmological constant.  

\subsubsection{Schwarzschild mass spectrum}

We first focus on the Schwarzschild system with $L_{\Lambda}\rightarrow +\infty$. Consider the Casimir generator
\be
\label{cazz1}
\T_0 = - V_1^2 P^2_1 = -  \frac{1}{4} \left( B P_B - A P_A\right)^2
\ee
which depends explicitly on the parameter $\epsilon$. Using the hamiltonian (\ref{ham0}) or (\ref{ham}) for $L_{\Lambda} \rightarrow +\infty$, it is direct to check that $\{ \T_0, H\} =0$, showing that $\T_0$ is a strong Dirac observable of the system. Computing its value for the Schwarzschild solution, one obtains
\be
\label{cazz2}
\T_0 = - \frac{L^2_0 L^2_{M}}{4 L^4_P}
\ee
which corresponds to the (rescaled) squared mass of the Schwarzschild black hole. The Schwarzschild mass spectrum is therefore given by the eigenvalues of the $\sl(2,\mathbb{R})$ Casimir. Since it is negative, the Casimir selects either the continuous serie or the complementary serie of the unitary irreducible representation of   $\sl(2,\mathbb{R})$. In the first case, the mass inherits a continuous spectrum given by
\begin{align}
\frac{L^2_0 L^2_M}{4L^4_{P}} =   s^2 + \frac{1}{4}  \;, \qquad \text{with}  \qquad s \in \mathbb{R}^{+}
\end{align}
This spectrum holds both for the exterior and interior regions. It is worth pointing that the continuity of this mass spectrum derived from purely symmetry-based arguments is in conflict with several other investigations on the quantization of the Schwarzschild black hole where the mass spectrum was postulated to be discrete \cite{Bekenstein:1995ju}.  Let us now investigate the role of $\T_0$ when $L_{\Lambda}$ remains finite. 

\subsubsection{Generating shift of the cosmological constant }

\label{sec3.5.2}

Consider now the general Schwarzschild-(A)dS system where $L_{\Lambda}$ remains finite. The Casimir $\T_0$ is no longer a Dirac observable since
\be
\{ \T_0, H \} = \frac{2\kappa L_P }{L^2_{\Lambda}} V^2_1 P_1
\ee
and it does not coincide anymore with the mass of the Schwarzschild-(A)dS spacetime given by (\ref{obs1}). It is  instructive to consider its square root , i.e. $S =  \sqrt{-\T_0} =V_1 P_1$, and compute its equations of motion which read
\begin{align}
\begin{aligned}
\label{flow}
\dot{S}  & =  -  \frac{\kappa L_P}{L^2_{\Lambda}} V_1\;, \qquad \ddot{S}   %=  -  \frac{1}{\tilde{L}_{\Lambda}} \{ V_1, H\} 
= - \frac{D}{L^2_{\Lambda}}  \;,  \qquad \dddot{S}   %=  - \frac{L^2_P}{L^2_{\Lambda}} \{ D, H\} 
= - \frac{\O_2}{\kappa L_P L^2_{\Lambda}} \;, \qquad \ddddot{S}   = 0
\end{aligned}
\end{align}
such that $\dddot{S}$ coincides with the (rescaled) Dirac observable $\O_2$. Integrating the above equations, we obtain
\be
\label{s}
S(\tau) = \delta + \gamma \tau + \frac{\beta}{2} \tau^2 - \frac{\O_2}{6 \kappa L_P L^2_{\Lambda}} \tau^3  
\ee
where $(\delta, \gamma, \beta)$ are integration constants. In order to fix them, one can compare to the Schwarzschild-(A)dS solution which gives
\be
 \beta = \gamma = 0 \;, \qquad \delta = \frac{L_0L_M}{2L^2_P}
\ee
We can now investigate the action of the generator $S= V_1P_1$ on the observables of the system and the flows it generates on the phase space. First, the bracket (\ref{flow}) provides the infinitesimal action of S on the hamiltonian, namely
\be
\delta_{\lambda} H = \lambda \{  S, H\}  = - \lambda \frac{\kappa L_P}{L^2_{\Lambda}} V_1
\ee
where $\lambda \in \mathbb{R}$ parametrizes the flow. It follows that the generator $S=V_1P_1$ induces a one-parameter family of deformation of the dynamics labeled by the cosmological constant. 
It allows one to modify the dynamics by shifting, and thus possibly removing, the interaction term associated to the cosmological constant. %Notice however that starting with a vanishing cosmological constant, the above bracket vanishes and $S$ becomes a Dirac observable. Therefore, the flow generated by $S$ is solution generating only if one starts with a non-vanishing cosmological constant. In the next section, we shall see how this point can be generalized to obtain a solution-generating map starting from $\Lambda=0$. 
Now, let us compute how the two Dirac observables $\O_1$ and $\O_2$ introduced in (\ref{obs1}) and (\ref{obs2}) evolve along the flow generated by $S$. Their evolution along that flow read
\begin{align}
\delta_{\lambda} \O_1 & = \lambda \{ S, \O_1 \} = \lambda \epsilon \frac{2}{3} \frac{\kappa L_P }{L^2_{\Lambda}} \frac{V_1}{P_2}  \\
\delta_{\lambda} \O_2 & = \lambda \{ S, \O_2 \} =  - \lambda \O_2
\end{align}
Therefore, the observable $\O_2$ is rescaled while $\O_1$ is corrected by a term which shifts its contribution involving the cosmological constant. Exponentiating the generator $S$, we obtain therefore the generator of finite shifts along the cosmological constant.  At the phase space level, the flow generated by $S=V_1P_1$ reflects the transformation of the action under the generalized translation (\ref{sym11}) which shifts the cosmological constant, i.e. when $\dddot{h}$ is a given constant.

We conclude that when $L_{\Lambda}$ is finite, the  $\sl(2,\mathbb{R})$ Casimir plays the role of shifting the cosmological constant. This  provides a new solution-generating technique which allows one to flow between the three different sectors of solutions of the Schwarzschild-(A)dS system. In the next section, we shall apply this solution-generating method and show how the Schwarzschild-(A)dS solution can be obtained from the pure Schwarzschild one through such finite transformation.

\section{M\"{o}bius covariance, trajectories mapping and solution-generating flow }

\label{sec4}

In this last section, we would like to understand how the M\"{o}bius symmetry (\ref{sym1}) acts at the level of the solutions. Being a physical symmetry, it should transforms gauge-inequivalent solutions to the field equations onto each other. More concretely, we expect that this symmetry connects Schwarzschild-(A)dS blakc hole with different parameters $(L_{M}, L_{\Lambda})$. Moreover,  the Schwarzschild-(A)dS geometry should be covariant under the newly found $\SL(2,\mathbb{R})\ltimes \mathbb{R}^3$ transformations. In this section, we demonstrate this covariance property explicitly and give the mapping between the trajectory parameters. Moreover, we show that one can use the translational sector to exhibit finite transformations which are not symmetry but stands as solution-generating transformation for the cosmological constant, allowing one to freely shift the value of this key parameter.

%In this section, we show how the symmetries uncovered in the previous section act on the physical trajectories, i.e. on the Schwarzschild-(A)dS solution.
%Concretely, we shall show that the Schwarzschild solution and its Schwarzschild-(A)dS extension are both covariant under the conformal reparametrization introduced previously. Moreover, we shall use the solution-generating transformations introduced earlier to construct the Schwarzschild-(A)dS solution from the pure Schwarzschild one, demonstrating thus the solution-generating status of the transformations (\ref{sym11} -\ref{sym13}).
%%%
\subsection{M\"{o}bius covariance of the Schwarzschild solution}
%\subsection{Trajectories at  $\Lambda=0$ and the Mo\"ebius symmetry under conformal reparametrizations}
%%%

Let us first fccus on the Schwarzschild solution. Considering the M\"{o}bius reparametrization of the $\tau$-coordinate, 
%\begin{align}
%\label{traj0}
%A^2 =  - \epsilon \frac{\mathcal{C}_1}{\mathcal{C}^2_2}\frac{\tau - \tau_1}{\tau - \tau_0} \;, \qquad B= \mathcal{C}_2(\tau-\tau_{0})\,,
%\end{align}
%where $(\tau_{0}, \tau_1)$ are two constants of integration and $( \mathcal{C}_1, \mathcal{C}_2)$ are two constants of motion. The relation to the standard Schwarzschild metric has been discussed in Section~\ref{sec1.2}. We recall that $A^2(\tau) = - \epsilon g^{\tau\tau}(\tau)$ and that $B^2(\tau) = g_{\theta\theta}(\tau)$. 
%We would like to understand how the M\"{o}bius symmetry of the action acts at the level of the solutions. For completeness, let us rewrite the symmetry transformations uncovered above. They can be expressed as conformal reparametrizations of the proper time $\tau$ where the fields transform as
%\begin{align}
%\tau\,\mapsto\, \ttau &=f(\tau)\,, \\
%\label{trans1}
%B(\tau)
%\,\mapsto\,
%\tB(\ttau) &=\dot{f}(\tau)^{\Delta_B}B(\tau) \;, \\
%\label{trans2}
%A(\tau)
%\,\mapsto\,
%\tA(\ttau)&=\dot{f}(\tau)^{\Delta_A}A(\tau)
%\,,
%\end{align}
%where the reparametrization function $f(\tau)$ corresponds to M\"{o}bius transformation such that
%\be
%f(\tau) = \frac{a\tau + b}{c\tau +b} \;, \;\;\; ad-bc =1\;, \qquad \dot{f} (\tau)= \frac{1}{(c\tau +d)^2}
%\ee
%and $\Delta_A$ and $\Delta_B$ are the conformal weights of the fields.
the two coordinates before and after the transformation are related through
\be
\tilde{\tau} = \frac{a\tau+b}{c\tau+d} \;, \qquad \tau = - \frac{d\ttau-b}{c\ttau-a}\,,
\ee
%The exponents $\Delta_A\alpha$ and $\beta$ are the dimensions of the fields $A$ and $B$ under conformal reparametrizations. We will see below that $\alpha=0$ and $\beta=\f12$, matching with  the results obtained in the previous works \cite{Geiller:2020xze}.
%
%More precisely, a straightforward calculation, done explicitly below, shows that the space of trajectories \eqref{traj0} is invariant under the $\SL(2,\R)$ subgroup of 
%conformal reparametrizations defined by Mo\"ebius transformations:
%\begin{align}
%f(\tau) & =\f{a\tau+b}{c\tau+d}=\ttau\,\qquad\textrm{with}\quad ad-bc=1
%\,,\\
%h(\tau) =\dot{f}(\tau) & =\f1{(c\tau+d)^{2}}=(-c\ttau+a)^{2}\,,\nn \\
%\tau & = - \f{d\ttau-b}{c\ttau-a}\,,
%\end{align}
%Let us turn to the conformal mapping of the fields. 
Now, consider the classical solutions for the $A$-field and $B$-field given by (\ref{profield}) expressed in the time $\tilde{\tau}$ and with the constants $(\tilde{\tau}_0, \tilde{\tau}_1, \tilde{\mathcal{C}}_1, \tilde{\mathcal{C}}_2)$. Explicitly, the $B$-field transforms as
%Now, consider that the transformed fields $\tA(\ttau)$ and $\tB(\ttau)$ are both classical solutions in the time $\ttau$ with constants of motion $\tka$, $\tR$, $\ttau_{0}$ and $\ttau_{1}$. Let us first look at the transformation law of the $B$-field:
\begin{align}
\begin{aligned}
\tB(\ttau)&= \tilde{\mathcal{C}}_2 (\ttau-\ttau_{0}) \\
& =\tilde{\mathcal{C}}_2 \left(\f{a\tau+b}{c\tau+d}-\ttau_{0}\right) \\
&= \frac{\tilde{\mathcal{C}}_2 (a-c\ttau_{0})}{c\tau +d}\left(\tau + \f{d\ttau_{0}-b}{c\ttau_{0}- a}\right)
=\dot{f}(\tau)^{\f12}\,\mathcal{C}_2(\tau-\tau_{0})=\dot{f}(\tau)^{\f12}\,B(\tau)\,,
\end{aligned}
\end{align}
which corresponds to (\ref{sym1}) as expected. This transformation shows that the solution for the $B$-field is covariant and that the involved constants $(\tilde{\tau}_0, \tilde{\mathcal{C}}_2)$ transform as follows
\begin{align}
\label{c2}
%\left\{
%    \begin{array}{ll}
%        \mathcal{C}_2 =  (a-c\ttau_{0}) \tilde{\mathcal{C}}_2  \\
%        \tau_{0}=- \f{d\ttau_{0}-b}{c\ttau_{0}- a} 
%    \end{array}
%\right.
%\qquad \qquad
\left\{
    \begin{array}{ll}
        \tilde{\mathcal{C}}_2 =  \dot{f}(\tau_0)^{-1/2} \mathcal{C}_2 \\
        \tilde{\tau}_{0} = f(\tau_{0})
    \end{array}
\right.
\end{align}
%This confirms the conformal dimension of the $B$ field as $\beta=\f12$.
%
Let us now look at the transformation law of the $A$-field, which should fix the mapping for the other two constants of motion $(\tau_1, \mathcal{C}_1)$. It can easily been shown that
\begin{align}
\tilde{A}^{2}(\tilde{\tau}) %& = - \epsilon\;  \frac{\tilde{\mathcal{C}}_1}{\tilde{\mathcal{C}}^2_2} \; \frac{\tilde{\tau} - \tilde{\tau}_1}{\tilde{\tau} - \tilde{\tau}_0} \\
%& = - \epsilon\;  \frac{\tilde{\mathcal{C}}_1}{\tilde{\mathcal{C}}^2_2} \; \left[ \frac{a\tau +b}{c\tau+d} - \tilde{\tau}_1\right] \left[  \frac{a\tau+b}{ c\tau +d} - \tilde{\tau}_0\right]^{-1} \\
%& = - \epsilon\;  \frac{\tilde{\mathcal{C}}_1 (a - c \tilde{\tau}_1)(a - c \tilde{\tau}_0)}{\tilde{\mathcal{C}}^2_2 (a - c \tilde{\tau}_0)^2} \; \left[ \tau + \frac{d \tilde{\tau}_1 -b}{c\tilde{\tau}_1-a} \right] \left[  \tau + \frac{d \tilde{\tau}_0 -b}{c\tilde{\tau}_0-a} \right]^{-1} \\
%& = - \epsilon \;  \frac{\mathcal{C}_1}{\mathcal{C}^2_2} \; \frac{ \tau - \tau_1}{\tau - \tau_0} \\
& = A^2(\tau)
\end{align}
which also fits with (\ref{sym1}). The transformations of the constants $(\tau_1,  \mathcal{C}_1)$ are given 
\begin{align}
\label{c1}
%\left\{
   % \begin{array}{ll}
  %      \mathcal{C}_1  & =(a-c\ttau_{0})(a-c\ttau_{1}) \;  \tilde{\mathcal{C}}_1 \\
 %       \tau_{1} &=- \f{d\ttau_{1}-b}{c\ttau_{1}- a} 
 %   \end{array}
%\right.
%\qquad \qquad
\left\{
    \begin{array}{ll}
        \tilde{\mathcal{C}}_1 & = \dot{f}^{-1/2}(\tau_{0}) \dot{f}^{-1/2}(\tau_{1})  \mathcal{C}_1\\
        \ttau_{1} & =f(\tau_{1})\,.
    \end{array}
\right.
\end{align}
The above transformation shows that the space of trajectories is indeed invariant under the $\SL(2,\mathbb{R})$ symmetry and that the Schwarzschild solution is covariant under such M\"{o}bius reparametrization. As expected, this physical symmetry maps the Schwarzschild solution with mass $\tilde{L}_M$, parametrized by the coordinate $\tilde{\tau}$,  into another Schwarzschild solution with mass $L_M$, parametrized by the coordinate $\tau$. Under this mapping, the mass parameter changes as 
\be
\tilde{L}_M = \tilde{ \tau}_1 - \tilde{\tau}_0  \;, \qquad \Rightarrow \qquad L_{M} =  \f{ad-bc}{(c\ttau_{1}- a)(c\ttau_{0}- a) } (\tilde{\tau}_1 - \tilde{\tau}_0) 
\ee
It shows that our symmetry transformation is not a diffeomorphism but connects the Schwarzschild geometries at different masses.
It follows that there is an equivalence class of Schwarzschild geometries related by this $\SL(2,\mathbb{R})$  symmetry transformation. This provides the first result of this section.

Before going further, it is interesting to discuss the fate of the boundary under our symmetry transformation. In our set-up, the boundary appears through the presence of the cut-off scales $(L_0, L_s)$ which are equivalent to $(\mathcal{C}_1, \mathcal{C}_2)$. These scales play the role of IR cut-off which fix the size of the region of interest in our symmetry-reduced model. Although these constants do not enter explicitly in the physical solution, and can be removed by a rescaling of the coordinates, they actually transform non trivially under our transformation (as (\ref{c1}) and (\ref{c2})). Therefore, the symmetry also acts on these scales, and thus on the implicit boundary that we have fixed by hand from the beginning. 

Now, in order to generalize this result to the Schwarzschild-(A)dS solution, we will have to consider more general maps acting on the space of trajectories and study their composition. To that end, it is useful to drop the tilde notation and locate explicitly the action of the conformal reparametrization. So re-establishing the explicit dependence of the trajectories on the constants of motion and constants of integration, i.e. writing $A[\tau_{0},\tau_{1}, \mathcal{C}_1, \mathcal{C}_2](\tau)$ and $B[\tau_{0},\tau_{1}, \mathcal{C}_1, \mathcal{C}_2](\tau)$ for the classical solutions,  we denote the M\"{o}bius mapping by $\cD_{f}$ and write:
\begin{align}
\begin{aligned}
\cD_{f}\Big{[}A[\tau_{0},\tau_{1}, \mathcal{C}_1, \mathcal{C}_2]\Big{]}(\tau)&=
A[f\triangleright(\tau_{0},\tau_{1}, \mathcal{C}_1, \mathcal{C}_2)](f(\tau))=A[\tau_{0},\tau_{1}, \mathcal{C}_1, \mathcal{C}_2](\tau)\,,\qquad
\\
\cD_{f}\Big{[}B[\tau_{0},\tau_{1}, \mathcal{C}_1, \mathcal{C}_2]\Big{]}(\tau)&=
B[f\triangleright(\tau_{0},\tau_{1}, \mathcal{C}_1, \mathcal{C}_2)](f(\tau))=\dot{f}(\tau)^{\f12}\,B[\tau_{0},\tau_{1}, \mathcal{C}_1, \mathcal{C}_2](\tau)\,,
\end{aligned}
\end{align}
%\beq
%&&A[f\triangleright(\ka,R,\tau_{0},\tau_{1})](f(\tau))=A[\ka,R,\tau_{0},\tau_{1}](\tau)\,,\qquad
%\\
%&&B[f\triangleright(\ka,R,\tau_{0},\tau_{1})](f(\tau))=\dot{f}(\tau)^{\f12}\,B[\ka,R,\tau_{0},\tau_{1}](\tau)\,,\nn
%\eeq
with the Mo\"ebius transformation acting directly on the trajectory parameters as:
\be
\label{eq:Dfparammap}
f\triangleright(\tau_{0},\tau_{1}, \mathcal{C}_1, \mathcal{C}_2)=(f(\tau_{0}),\,f(\tau_{1}), \dot{f}(\tau_{0})^{-\f12}\dot{f}(\tau_{1})^{-\f12} \mathcal{C}_1 ,\,\dot{f}(\tau_{0})^{-\f12} \mathcal{C}_2)
\,.
\ee
This provides a compact notation of the above mapping. 
We can now turn to the generalization of this mapping to 
%This shows that these constitute symmetries of the theory for a vanishing cosmological constant $\Lambda=0$.
%
%Let us keep in mind that this relation only holds if $f$ is a Mo\"ebius transformation. If $f$ is an arbitrary function, then one can show that the Schwarzian derivative will appear in this mapping and will also extend the space of classical trajectories \cite{Geiller:2020xze}. This leads to an action of the Virasoro group (with central charge) on the theory.
  to the Schwarzschild-(A)dS case. To that end, it will be useful to first discuss the fate of the translational symmetry on the Schwarzschild-(A)dS solution.

%%%
\subsection{Conformal bridge: From Schwarzschild to Schwarzschild-(A)dS}
%\subsection{Translating $\Lambda=0$ to $\Lambda\ne0$ and Schwarzschild-(A)dS automorphisms}
%\subsection{Translations from $\Lambda=0$ to $\Lambda\ne0$ and Schwarzschild-(A)dS automorphisms}
%%%
Consider therefore the Schwarzschild-(A)dS solutions for the $A$-field and $B$-field given by (\ref{profield})
%\be
%A^2 = \epsilon \left[ \frac{(\tau - \tau_0)^2}{3L^2_{\Lambda}}  - \frac{\mathcal{C}_1}{\mathcal{C}^2_2}\frac{\tau - \tau_1}{\tau - \tau_0} \right] \;, \qquad B= \mathcal{C}_2(\tau-\tau_{0})\,,
%\ee
and let us focus on the transformations (\ref{sym11})
%\begin{align}
%\label{eqn:translation}
%\tau
%\mapsto
%\ttau & =\tau\;, \\
%\qquad
%B(\tau)
%\,\mapsto\,
%\tB(\tau) & =B(\tau)\;, \\
%\qquad
%A(\tau)^{2}
%\,\mapsto\,
%\tA(\tau)^{2}& =A(\tau)^{2}+2h \f{\dot{B}}{B}-\dot{h}
%\,,
%\end{align}
where $\ddddot{h}=0$. This transformation is not a symmetry but generates instead a shift of the action which mimics a cosmological constant. See the second term in (\ref{varT}) for the explicit transformation of the action. In the following, we shall therefore consider this case such that
\begin{align}
h(\tau) & := \alpha \tau^3 + \beta \tau^2 + \gamma \tau + \delta 
%& : = \alpha  (\tau-\tau_0)^3 + (\beta + 3 \alpha \tau_0) (\tau - \tau_0)^2 + (\gamma + 2 \beta \tau_0 +3 \alpha \tau^2_0) (\tau - \tau_0) + \delta + \gamma \tau_0 + \beta \tau^2_0 + \alpha \tau^3_0
\end{align}
where $\dddot{h} = 6 \alpha$ is a constant. This fits with the generator $S= V_1 P_1$ given in (\ref{s}) which we have identified with the generator of deformation parametrized by the cosmological constant.
Now let us translate the mapping $(A,B)\mapsto (\tA,\tB)$ given above into a mapping between the constants of motion parametrizing the classical trajectories. First, since translations do not change the $B$-field, one has
\be
B=\mathcal{C}_2(\tau-\tau_{0})= \tilde{\mathcal{C}}_2(\tau-\ttau_{0})=\tB\,,
\ee
it means that translations do not shift neither the $B$-velocity $\mathcal{C}_2 =\tilde{\mathcal{C}}_2$, nor the singularity time $\tau_{0}=\ttau_{0}$. The non-trivial part is the transformation law for the $A$-field which provides the transformations of the constants $(\tau_1, \mathcal{C}_1, L_{\Lambda})$. It reads:
\begin{align}
\begin{aligned}
\tA^{2} & =  \epsilon \left[ \frac{(\tau - \tau_0)^2}{3\tilde{L}^2_{\Lambda}}  - \frac{\tilde{\mathcal{C}}_1}{\mathcal{C}^2_2}\frac{\tau - \tilde{\tau}_1}{\tau - \tau_0} \right] \\
& = \epsilon \left[ \frac{(\tau - \tau_0)^2}{3 L^2_{\Lambda}}  - \frac{\mathcal{C}_1}{\mathcal{C}^2_2}\frac{\tau - \tau_1}{\tau - \tau_0} \right] + 2h \frac{\dot{B}}{B} - \dot{h}\\
& = \epsilon \left[   \left( \frac{1}{L^2_{\Lambda}} -  3\epsilon \alpha \right) \frac{(\tau - \tau_0)^2}{3} - \frac{K_1 \tau - K_2 }{\mathcal{C}^2_2 (\tau- \tau_0)}  \right] \\
& = A^2
\end{aligned}
\end{align}
%\begin{align}
%=
%\f{\tLambda}3(\tau-\tau_{0})^{2}-\f{\tka(\tau-\ttau_{1})}{R^{2}(\tau-\tau_{0})}
%=
%\f{\Lambda}3(\tau-\tau_{0})^{2}-\f{\ka(\tau-\tau_{1})}{R^{2}(\tau-\tau_{0})}
%+
%2\phi\f{\dot{B}}{B}-\dot{\phi}
%=
%A^{2}
%\,,
%\end{align}
where 
\be
K_1 = \mathcal{C}_1 - \epsilon \mathcal{C}^2_2 (\gamma + 2 \beta \tau_0 + 3 \alpha \tau^2_0) \;, \qquad K_2 =  \mathcal{C}_1 \tau_1 + \epsilon  \mathcal{C}_2^2 (2\delta - \gamma \tau_0 - \alpha \tau^3_0)
\ee
This result shows that the Schwarzschild-(A)dS solution for the $A$-field is also covariant under this specific translation.
It is then straigthforward to read the transformation of the constants $(\tau_1, \mathcal{C}_1, L_{\Lambda})$. Introducing the standard notation $\Lambda = L^{-2}_{\Lambda}$ for the cosmological constant, one obtains
\begin{align}
\begin{aligned}
\label{eqn:mappedparam}
\tLambda & =\Lambda-3\epsilon \alpha
\,, \\ 
\tilde{\mathcal{C}_1} & = \mathcal{C}_1 - \epsilon \mathcal{C}_2^{2}(\gamma+2\beta\tau_{0}+3\alpha\tau_{0}^{2})
\,,\\
\tilde{\mathcal{C}}_1 \tilde{\tau}_{1} & = \mathcal{C}_1\tau_{1} + \epsilon \mathcal{C}^2_2(2\delta +\gamma \tau_{0}-\alpha \tau_{0}^{3})
\,.
\end{aligned}
\end{align}
%This confirms the mapping derived from directly applying the translation transformation to the action principle: when $\tLambda=0$ vanishes, the cosmological constant for the $A$ and $B$ fields is entirely given by the third derivative of the translation parameter, $\Lambda=3u=\tfrac12\phi^{(3)}$.
%
%
Having obtained the mapping between the constants involved in the physical trajectory, it is again convenient to drop the tilde notation for the transformation and introduce a specific notation for the translation mapping as $\cT_{\phi}$ acting as:
\begin{align}
\begin{aligned}
\cT_{\phi} \Big{[}A[\tau_{0},\tau_{1}, \mathcal{C}_1, \mathcal{C}_2, \Lambda]\Big{]}^{2} 
&= A[\phi\triangleright(\tau_{0},\tau_{1}, \mathcal{C}_1, \mathcal{C}_2, \Lambda)]^{2}\\
& = A[\tau_{0},\tau_{1}, \mathcal{C}_1, \mathcal{C}_2, \Lambda]^{2}
+
2h \dot{B}[\tau_{0},\tau_{1}, \mathcal{C}_1, \mathcal{C}_2, \Lambda] B^{-1}[\tau_{0},\tau_{1}, \mathcal{C}_1, \mathcal{C}_2, \Lambda]-\dot{h}
\,,
%\cT_{\phi}\Big{[}B[\tau_{0},\tau_{1}, \mathcal{C}_1, \mathcal{C}_2, \Lambda]\Big{]}&=&
%B[\phi\triangleright(\tau_{0},\tau_{1}, \mathcal{C}_1, \mathcal{C}_2, \Lambda)]\,,
\end{aligned}
\end{align}
%\beq
%&&A[f\triangleright(\ka,R,\tau_{0},\tau_{1})](f(\tau))=A[\ka,R,\tau_{0},\tau_{1}](\tau)\,,\qquad
%\\
%&&B[f\triangleright(\ka,R,\tau_{0},\tau_{1})](f(\tau))=\dot{f}(\tau)^{\f12}\,B[\ka,R,\tau_{0},\tau_{1}](\tau)\,,\nn
%\eeq
with the translation acting  on the trajectory's parameter multiplet $(\tau_{0},\tau_{1}, \mathcal{C}_1, \mathcal{C}_2, \Lambda)$ as:
\begin{align}
& h \triangleright \mathcal{C}_1  = \mathcal{C}_1 - \epsilon \;  \mathcal{C}_2^{2}\dot{h}(\tau_{0}) \,,\qquad \qquad \qquad \qquad \qquad  h \triangleright \mathcal{C}_2   = \mathcal{C}_2 \,, \\
& h \triangleright \tau_{1} =\frac{\tau_1 + \epsilon \;  \mathcal{C}_1^{-1}  \mathcal{C}_2^2 (2h(\tau_0) - \tau_0 \dot{h}(\tau_0))}{1 - \epsilon \;  \mathcal{C}_1^{-1}  \mathcal{C}_2^2 \dot{h}(\tau_0)} \,, \qquad  \;\;\;\; h \triangleright \tau_{0} = \tau_{0}\,,
\end{align}
while 
\begin{align}
\label{cos}
h \triangleright \Lambda = \Lambda - \frac{\dddot{h} (\tau_{0}) }{2}\;, 
\end{align}
%Notice that $h(\tau)$ is a third order polynomial so that its third derivative is constant, $\dddot{h}(\tau)= \dddot{h} (\tau_{0})$. 
The transformation  (\ref{cos}) of the cosmological constant shows that these translations allow one to turn the cosmological constant $\Lambda$ on and off, and even shift between arbitrary (positive and negative) values of $\Lambda$, thereby defining mapping between the asymptotically dS, AdS and flat Schwarzschild solutions. These transformations are therefore not symmetry of the action but stand as solution-generating maps which connect the Schwarzschild and Schwarzschild-(A)dS solutions.

Now, the specific translation discussed here allows one to freely shift the value of the cosmological constant. Starting from the Schwarzschild-(A)dS solution, it is therefore possible to perform such transformation to remove the cosmological constant and map it to the pure Schwarzschild solution. One can then perform a M\"{o}bius reparametrization and finally perform a second translation which sets back the cosmological constant to its original value. It follows from this three-steps-transformation that the Schwarzschild-(A)dS solution is also covariant under M\"{o}bius transformation (suitably composed with specific translations). The explicit proof of this last step can be found in Appendix~\ref{appC}.
%Now, let us see how this new map allows one to generalize the M\"{o}bius covariance of the Schwarzschild solution to the Schwarzschild-(A)dS one.

%%%

\section{Discussion}

\label{sec5}

In this work, we have shown that the Schwarzschild-(A)dS black hole mechanics enjoys a hidden Noether symmetry under the group SL$(2,\mathbb{R})\ltimes \mathbb{R}^3$. This symmetry is realized only after the gauge invariance of the symmetry-reduced action (\ref{gfac}) is fixed. Being a physical symmetry, it maps gauge-inequivalent Schwarzschild-(A)dS solutions with different mass and cosmological constant onto each other. A direct consequence of this hidden symmetry is that one can associate a set of non-trivial charges with the Schwarzschild-(A)dS black hole whose algebra fully dictates the underlying geometry on both side of the horizon(s). This result generalizes the symmetry structure found for the Schwarzschild black hole interior mechanics  in \cite{Geiller:2020xze} and in isotropic cosmological systems in \cite{BenAchour:2019ufa, BenAchour:2020xif, Achour:2021lqq, BenAchour:2020ewm, BenAchour:2020njq}. 

This generalization has been obtained by considering the Kantowski-Sachs symmetry-reduced homogeneous model of GR with a cosmological constant which admits the family of Schwarzschild-(A)dS black hole as solutions. The results obtained in this work are the following:
\begin{itemize}
\item By considering the metric (\ref{ansatz}) with parameter $\epsilon =\pm1$, we have shown that one can treat at once both the $T$-region ($\epsilon = +1$) and $R$-region ($\epsilon=-1$) of the underlying Schwarzschild-(A)dS geometry. This set-up allows one to foliate the $T$-region with spacelike hypersurface and $R$-region with timelike ones, and switch between each other by flipping the sign of the parameter $\epsilon$. The SL$(2,\mathbb{R})\ltimes \mathbb{R}^3$ symmetry uncovered in this work acts therefore on fields living on two distinct hypersurfaces depending on the region of interest. It follows that in the exterior region, one can associate a set of non-trivial charges to any time-like hypersurface at a given radius $\tau$. While the expression of the charge is $\tau$-dependent, the charge algebra is the same at any value of the radial coordinate, such that the symmetry is realized in the same way near the horizon as well as at large radius. The same is true in the interior region where the $\tau$-coordinate is now a time coordinate labelling spacelike hypersurface. This extension of the result found in \cite{Geiller:2020xze} for the Schwarzschild interior mechanics provides the first main result of this work. 
\item Moreover, we have shown that when turning on the cosmological constant, the SL$(2,\mathbb{R})$ sector of the symmetry has to be modified. The new symmetry transformation are given by (\ref{sym1}-\ref{sym3}) where the corrective term affects only the $A$-field through (\ref{corr}). These new transformations find an elegant interpretation in term of the conformal bridge we have identified.  Indeed, by noticing that one can freely change the value of the cosmological constant with suitable translations (which are not symmetries of the action), one can start from the Schwarzschild-(A)dS reduced action, acts with the conformal bridge to remove the cosmological constant, perform the M\"{o}bius transformation identified in \cite{Geiller:2020xze} for the pure Schwarzschild case, and finally acts once more with the conformal bridge map to set back the cosmological constant to its original value.  This process consists in a composition of a M\"{o}bius transformation with suitable translations which can be compactly written as (\ref{compo}). From that perspective, the conformal bridge identified in this work plays a key role in the realization of the SL$(2,\mathbb{R})$ symmetry of the Schwarzschild-(A)dS black hole mechanics. The conserved charges generating this hidden symmetry are given in (\ref{ch1}-\ref{ch3}) and (\ref{cht1}-\ref{cht3})
\item At the hamiltonian level, this symmetry is reflected in an extended version of the CVH algebra first identified in cosmological models in \cite{BenAchour:2018jwq, BenAchour:2019ywl, BenAchour:2017qpb}. Just as for the Schwarzschild case, this algebra can be shown to be isomorphic to the three dimensional Poincar\'e algebra $\sl(2,\mathbb{R})\ltimes \mathbb{R}^3$ and can be used to rewrite the conserved charges in an intuitive form given by (\ref{c1} - \ref{c6}). An interesting outcome of our analysis is that, although the charges depend explicitly on the parameter $\epsilon$ which distinguishes between the $T$-region and $R$-region, the charge algebra does not dependent on this parameter. It follows that the same algebraic structure is at play in each region, although acting on fields living on different hypersurfaces. Physically, these conserved charges can be used to reconstruct the Schwarzschild-(A)dS solution and their algebra fully dictates the geometry on both side of the horizons (the black hole or/and the cosmological ones). An interesting question is whether one can view the matching of the charges at the horizon as an algebraic realization of the standard junction condition in GR ? We leave this intriguing question for future work.
\item Another interesting outcome of this investigation concerns the role played by the invariant Casimirs of the $\sl(2,\mathbb{R})\ltimes \mathbb{R}^3$  Poincar\'e algebra. Among the three Casimirs, only the one of the  $\sl(2,\mathbb{R})$ algebra is non-vanishing and given by (\ref{cazz1}).  In the pure Schwarzschild case, it coincides with the squared mass, therefore labelling the equilibrium thermodynamical states of the black hole. When the cosmological constant is turned on, the $\sl(2,\mathbb{R})$ Casimir does not match with the mass of the black hole given by (\ref{obs1}), but play instead a new surprising role. As shown in Section~\ref{sec3.5.2}, it generates a one-parameter family of deformation whose parameter is nothing else than the cosmological constant. Therefore, the existence of such generator reveals a new map connecting the Schwarzschild, Schwarzschild-AdS and Schwarzschild-dS black hole solutions. Setting the mass to zero, one obtains a simple map between the Minkowski and the vacuum AdS or dS spacetimes. This give rise to a new conformal bridge which generalizes the one identified in cosmology which relates the flat FLRW model to its (A)dS extensions or to the $k=\pm1$ universes \cite{BenAchour:2020xif, Achour:2021lqq}. This provides the last key result of this work. The fact that one can map different black holes solutions with radically different asymptotic behavior is quite remarquable and begs for further investigations. Whether this conformal bridge or some extension of it can be of any use to connect results in AdS holography to its flat or dS version remain to be explored.
\end{itemize}

From a more general perspective, the present work provides one more example illustrating the emergence of hidden symmetries in homogeneous symmetry-reduced gravity \cite{BenAchour:2019ufa, BenAchour:2020xif, Achour:2021lqq, BenAchour:2020ewm, BenAchour:2020njq}. Nevertheless, the status of such hidden symmetry remains puzzling for several reasons. First, because of homogeneity, the boundary seems to not play any role in the emergence of this symmetry and the information on the boundary carried by the non-trivial conserved charges shows up only through the cut-off scales fixing the size of the fiducial cell, namely the scales $(L_0, L_s)$ (or equivalently $(\mathcal{C}_1, \mathcal{C}_2)$) introduced in Section~\ref{sec1}. This specific feature of homogeneous models, where the dynamics of fields on the boundary matches exactly the one in the bulk, contrast with the inhomogeneous case, in which the boundary play a central role in the construction of the conserved charges.  Second, the symmetry transformations that we consider cannot be represented as diffeomorphism of the metric, whose components transform as primary fields with a suitable (possibly different) conformal weight. Despite these key differences, the transformations exhibited in this work stand as well defined physical symmetries of the Schwarzschild-(A)dS black hole mechanics. We stress that our analysis holds for both the anti-de Sitter and de Sitter case.

At this stage, several questions remain to be addressed. First of all, what is the maximal symmetry group of the gauge-fixed system~? The three-dimensional Poincar\'e structure uncovered here and in \cite{Geiller:2020xze} might well be only the corner of a larger structure. Previous results obtained in cosmological systems have revealed that the CVH algebra belongs actually to a larger $\so(3,2)$ conformal algebra of observables \cite{BenAchour:2020njq}. Whether a similar extension exists for the black hole phase space remains to be investigated. Finally, can we use this symmetry structure to unravel physical properties of the Schwarzschild black hole~? Does this symmetry play a role in black hole perturbation theory, for example, by giving rise to new invariance for test fields propagating on it~? We expect to address these fascinating questions in the near future.

\section*{Acknowledgments}

The work of J. Ben Achour is supported by the Alexander von Humboldt foundation. 

%%%%%%%%%%%%%%%%%%%%%%%%%%%%%%%

 \appendix

\section{Schwarzschild-(A)dS solution and observables}

\label{appA}
In this appendix, we show that the mechanical action (\ref{acc}) admits indeed the Schwarzschild-(A)dS black hole geometry as a solution. While the derivation is textbook, it will be useful  to review it as it allows one to identify the constant of motions of our system. Varying w.r.t to the three fields $(\N, A, B)$, and introducing the proper $\tau$-coordinate $\rd \tau = \N \rd \eta$ to write down the equations, one obtains
\begin{align}
\label{eom1}
\E_{\N} & = \epsilon \left( \frac{1}{L^2_s} - \frac{B^2}{L^2_{\Lambda}}\right) +A^2 \dot{B}^2 + 2 A \dot{A} B \dot{B} \simeq 0 \\
\label{eom2}
\E_{A} & = \rd_{\tau} (A B \dot{B} ) - A \dot{B}^2 + B \dot{B} \dot{A} \simeq 0 \\
\label{eom3}
\E_{B} & = \rd_{\tau} ( A^2 \dot{B} + A \dot{A} B) - A\dot{A}\dot{B} - \frac{\epsilon}{L^2_{\Lambda}} B \simeq 0
\end{align}
where $\simeq$ refers to on-shell and a dot now refers to a derivative w.r.t the new coordinate $\tau$. Is is then straightforwrd to show that
\begin{align}
\E_A & = - AB  \; \rd^2_{\tau} B \simeq 0 \qquad \rd_{\tau} \left( \E_{\N} - \frac{\epsilon}{L^2_s}\right) = 2 \dot{B} \E_{B} - 2 \dot{A} \E_A \simeq 0
\end{align}
From these expressions, we identify the two constants of motion $\rd_{\tau} \mathcal{C}_1 = \rd_{\tau} \mathcal{C}_2 \simeq0$ which read
\begin{align}
 \mathcal{C}_1 =  - \epsilon \left[ A^2 \dot{B}^2 + 2 A \dot{A} B \dot{B } - \frac{\epsilon}{L^2_{\Lambda}} B^2  \right]\;, \qquad \mathcal{C}_2 =  \rd_{\tau} B  \;,
\end{align}
 such that $[\mathcal{C}_1] = L^{-2} $ and $[\mathcal{C}_2] = L^{-1}$. The interpretation of these constants of motion is straightforward : $\mathcal{C}_2$ is the velocity of the $B$-field, i.e the physical radius, while Eq~(\ref{eom1}) shows that $\mathcal{C}_1= L^{-2}_s$ which corresponds to the constant curvature of the 2-sphere.
The second observable $\mathcal{C}_2$ can be used to solve for the $B$-field which reads
\be
%\label{bfield}
B(\tau) = \mathcal{C}_2 \left( \tau-\tau_0 \right)
\ee
where $\tau_0$ is an integration constant with dimension of length scale. On shell, the second observable can be recast into
\be
\label{profA}
\mathcal{C}_1 =- \epsilon \left[  \rd_{\tau} (A^2 B \dot{B}) - \frac{\epsilon }{L^2_{\Lambda}} B^2 \right]
\ee
%It is then convenient to fix the constants to
%\be
%\frac{\mathcal{C}_1}{L^2_1} = - \frac{\epsilon}{L^2_s}\;, \qquad \frac{\mathcal{C}_2}{L_2} = \frac{1}{L_s}
%\ee
Integrating Eq~(\ref{profA}) for the $A$-field, one obtains
\begin{align}
%\label{afield}
A^2 = \epsilon \left[ \frac{(\tau - \tau_0)^2}{3L^2_{\Lambda}}  - \frac{\mathcal{C}_1}{\mathcal{C}^2_2}\frac{\tau - \tau_1}{\tau - \tau_0} \right]
\end{align}
where $\tau_1$ is a second constant of integration. In order to make contact with the standard Schwarzschild-(A)dS solution, one can use the invariance under translation of the system and apply the transformation $\tau \rightarrow \tau + \tau_0$. Introducing the length scale 
\be
L_{M} = \tau_1 - \tau_0
\ee
the $(A,B)$ fields read
\be
A^2(\tau) = - \epsilon \frac{\mathcal{C}_1}{\mathcal{C}^2_2} \left( 1 - \frac{L_{M}}{\tau} - \frac{\mathcal{C}^2_2}{3\mathcal{C}_1 L^2_{\Lambda}} \tau^2 \right) \;, \qquad B(x) = \mathcal{C}_2\tau
\ee
Now, the solution depends on two pairs of constants: the length scale parameters $(L_M, L_{\Lambda})$ and the two constants $(\mathcal{C}_1, \mathcal{C}_2)$. The former are actual parameters of the family of solutions while the latter have been introduced to fix the extension of the region of interest and avoid divergencies. As expected, they can be re-absorbed by performing the following rescalings:
\be
\label{rescaling}
\tau \rightarrow \frac{\mathcal{C}_2}{\sqrt{\mathcal{C}_1}} \tau \;, \qquad y \rightarrow \frac{\sqrt{\mathcal{C}_1}}{\mathcal{C}_2} y \;, \qquad L_{\Lambda} \rightarrow  \frac{\mathcal{C}_2}{\sqrt{\mathcal{C}_1}}L_{\Lambda} 
\ee
Then, using that $\mathcal{C}_1 = L^{-2}_s$, the solution line element (\ref{met}) written in proper coordinate $\rd \tau= \N \rd \eta$ becomes
\begin{align}
\label{metsol}
\rd s^2 = -  \left( 1 - \frac{L_{M}}{\tau} - \frac{\tau^2}{3L^2_{\Lambda}}\right) \rd y^2+  \left( 1 - \frac{L_{M}}{\tau} - \frac{\tau^2}{3L^2_{\Lambda}}\right)^{-1} \rd \tau^2  +  \tau^2 \rd\Omega^2
\end{align}
such that the physical solution depends only on the two parameters $(L_M, L_{\Lambda})$. The resulting metric corresponds to the standard form of the Schwarzschild-(A)dS metric where $L_{M}$ is the Schwarzschild mass. The Schwarzschild-dS background corresponds to $L_{\Lambda}>0$ while its Schwarzschild-AdS counterpart is obtained by the map $L_{\Lambda} \rightarrow i L_{\Lambda}$.

\section{Symmetry transformations and Noether charges computation}
\label{appB}

In this appendix, we present the detailed computation of variation of the action under the hidden symmetries identified in Section~\ref{sec2}. We introduce the dimensionless quantity 
\be
\kappa = \frac{L_0L^2_s}{L^3_P}
\ee
for simplicity. It encodes the ratio between the effective size of the system, i.e the IR cut-off, and the Planck volume, i.e the UV cut-off our our system.
%Consider thus the symmetry reduced action and let us gauge fix it by reabsorbing the lapse field by introducing the $\tau$-coordinate
%\be
%\label{tau}
%\rd \tau = \N \rd \eta
%\ee
The gauge fixed action reads
\begin{align}
%\label{gfac}
S_{\epsilon} [A, B] = \frac{L_0 L^2_s}{ \epsilon L^2_P}\int  \rd \tau \left[ \epsilon  \left( \frac{1}{L^2_s} - \frac{B^2}{L^2_{\Lambda}} \right)  - A^2 \dot{B}^2 - 2 AB \dot{B} \dot{A} \right]
\end{align}
Consider now the following transformations
\begin{align}
\begin{aligned}
\label{symm1}
\tau \rightarrow \tilde{\tau} & = f(\tau) \\
B(\tau) \rightarrow \tilde{B}(\tilde{\tau}) & = \dot{f}^{1/2} B(\tau) \\
A^2(\tau) \rightarrow \tilde{A}^2(\tilde{\tau}) & = A^2(\tau) + \chi(\tau) 
\end{aligned}
\end{align}
where $f(\tau)$ is an arbitrary function and where
\begin{align}
\label{corr}
\chi (\tau)&:= 2 h \frac{\dot{B}}{B} - \dot{h} - \frac{2 h\circ f}{\dot{f}} \left[ \frac{\dot{B}}{B} + \frac{\ddot{f}}{2\dot{f}}\right] + \dot{h} \circ f \;, \\
\label{h}
h (\tau) & :=\frac{\epsilon}{3L^2_{\Lambda}} \tau^3  \;,
%h_{f} (\tau) & := h \circ f (\tau) = \frac{\epsilon}{3L^2_{\Lambda}} f(\tau)^3
\end{align}
%These transformations are a generalization of the ones considered initially in symmetry-reduced cosmological models in \cite{BenAchour:2019ufa, BenAchour:2020xif, Achour:2021lqq} and extended to black hole interior midi-superspace in \cite{Geiller:2020xze}.
%When $L_{\Lambda} \rightarrow + \infty$, the field $A(\tau)$ remains invariant since $\chi(\tau) =0$ and one recovers the transformations introduced in \cite{Geiller:2020xze} for the pure Schwarzschild case. 
%As we shall see now, the additional term $\chi(\tau)$ in the transformation (\ref{sym3}) provides the required corrections to generalize the result of \cite{Geiller:2020xze} when turning on the cosmological constant. 
Under these transformations, the reduced action varies as
\begin{align}
\label{varaccc}
\Delta S_{\epsilon}  
& = \epsilon \kappa L_P \int \rd \tau  B^2 \left\{ \frac{1}{2} \text{Sch}[f]  \left [  A^2 + \chi - 4 \dot{h}\circ f \right]  + \frac{h\circ f}{\dot{f}}  \; \frac{\rd}{\rd \tau} \text{Sch}[f] \right\}  + \epsilon \kappa L_P \int \rd \tau \frac{\rd F}{\rd \tau} 
%& +  \epsilon c L_P \int \rd \tau B^2 \left\{  \frac{\epsilon}{L^2_{\Lambda}}  \left[ \left( \frac{\dddot{h}}{2} - 1\right) -  \dot{f}^2 \left(  \frac{\dddot{h}\circ f}{2} - 1\right) \right] \right\} \\
%& +  \epsilon c L_P \int \rd \tau \frac{\rd}{\rd \tau} \left\{ \frac{\epsilon}{L^2_s} ( f - \tau) + \left[  \frac{\rd^2}{\rd \tau^2} \left( h - \frac{h\circ f}{\dot{f}}\right) -\frac{\ddot{f}}{\dot{f}} (A^2 + \chi )   \right] \frac{B^2}{2} -   \left( h - \frac{h\circ f}{\dot{f}}\right) \dot{B}^2   \right\}
\end{align}
which splits in two terms. The first one encodes the conformal anomaly of the action variation which is governed by the Schwarzian derivative of the reparametrization function $f(\tau)$ defined as
\be
\label{sch}
\text{Sch}[f]  = \frac{\dddot{f}}{\dot{f}} - \frac{3}{2} \left( \frac{\ddot{f}}{\dot{f}}\right)^2
\ee
The second term is the total derivative term given by
\be
\label{F}
F = \left\{ \frac{\epsilon}{L^2_s} ( f - \tau) + \left[  \frac{\rd^2}{\rd \tau^2} \left( h - \frac{h\circ f}{\dot{f}}\right) -\frac{\ddot{f}}{\dot{f}} (A^2 + \chi )   \right] \frac{B^2}{2} -   \left( h - \frac{h\circ f}{\dot{f}}\right) \dot{B}^2   \right\}
\ee
Now, by definition of the Schwarzian, the anomalous contribution vanishes provided we restrict the reparametrization function $f(\tau)$ to a M\"{o}bius reparametrization, i.e.
\be
\label{f}
f(\tau) = \frac{a \tau + b}{c\tau + d} \;, \qquad ad-bc \neq 0
\ee
%while the second term vanishes since
%\be
%\ddddot{h} (\tau) = \frac{2\epsilon }{L^2_{\Lambda}} \qquad \text{such that} \qquad \ddddot{h} \circ f (\tau)= \frac{2\epsilon }{L^2_{\Lambda}}
%\ee
Therefore, for our specific choice (\ref{h}) and (\ref{f}), the variation of the action reduces to a total derivative term 
%\begin{align}
%\Delta S_{\epsilon}  
%& =  \epsilon c L_P \int \rd \tau \frac{\rd}{\rd \tau} \left\{ \frac{\epsilon}{L^2_s} ( f - \tau) + \left[  \frac{\rd^2}{\rd \tau^2} \left( h - \frac{h\circ f}{\dot{f}}\right) -\frac{\ddot{f}}{\dot{f}} (A^2 + \chi )   \right] \frac{B^2}{2} -   \left( h - \frac{h\circ f}{\dot{f}}\right) \dot{B}^2   \right\}
%\end{align}
which shows that the Schwarzschild-(A)dS mechanics enjoys indeed a Noether symmetry under the conformal reparametrization introduced above. Since this result is independent of the value of $\epsilon$, it implies that the symmetry acts both in the exterior and interior of the Schwarzschild-(A)dS black hole. We now derive the Noether charges generating this symmetry.

Consider thus an infinitesimal transformation of the form $f(\tau) = \tau + \xi(\tau)$. For an infinitesimal M\"{o}bius transformation, the function $\xi(\tau)$ corresponds to 
\begin{align}
\label{ep}
\xi(\tau) = \left\{
    \begin{array}{ll}
        \sigma  & \mbox{for translation }  \\
        \sigma \tau & \mbox{for dilatation} \\
        \sigma \tau^2  & \mbox{for special conformal transformation}
    \end{array}
\right.
\end{align}
which implies that $\xi(\tau)$ is at most quadratic in $\tau$ such that $\dddot{\xi}=0$. The infinitesimal transformations of the fields read
\begin{align}
\begin{aligned}
\delta \tau & = \tilde{\tau} - \tau  = \xi \\
\delta B & = \tilde{B} (\tau) - B(\tau)  = - \xi \dot{B} + \frac{1}{2} \dot{\xi} B \\
\label{da}
\delta ( A^2 ) & =  \tilde{A}^2 (\tau) - A^2(\tau) = - \xi \frac{\rd A^2}{\rd\tau} + \chi %= - \xi \frac{\rd A^2}{\rd\tau} + \chi
\end{aligned}
\end{align}
where the last term $\chi$ in (\ref{da}) is given, to first order in $\xi$, by the following expression
\begin{align}
\chi %& = \left( \dot{h}\circ f - \dot{h}\right) + 2 \left( h - \frac{h\circ f}{\dot{f}}\right) \frac{\dot{B}}{B} - \frac{h\circ f}{\dot{f}} \frac{\ddot{f}}{\dot{f}}\\
%& = \frac{\epsilon}{L^2_{\Lambda}} \left[ \left( \tau^2 + 2 \tau\xi - \tau^2 \right) + \frac{2\dot{B}}{B} \left( \frac{\tau^3}{3} - \frac{\tau^3}{3} - \tau^2 \xi + \frac{\tau^3}{3}  \dot{\xi}\right) - \frac{\tau^3}{3} \ddot{\xi}\right] + \O(\xi^2) \\
& =  \frac{2\epsilon}{L^2_{\Lambda}} \left[ \tau \xi - \frac{\tau^3}{6} \ddot{\xi}+ \left( \frac{\tau^3}{3} \dot{\xi} - \tau^2 \xi\right) \frac{\dot{B}}{B}\right] + \O(\xi^2)
\end{align}
Using these infinitesimal variations of the fields, the infinitesimal variation of the action can be written as
\begin{align}
\begin{aligned}
\delta S_{\epsilon} & = \epsilon \kappa L_{P}\int \rd \tau \frac{\rd}{\rd \tau} \left[ \xi \left( \frac{\epsilon}{L^2_{\Lambda}} B^2 + A^2 \dot{B}^2 + 2 A \dot{A}B \dot{B}\right) - \ddot{\xi} \frac{A^2B^2}{2}  \right. \\
& \left. \qquad \qquad \qquad \qquad \qquad \qquad - \frac{\epsilon}{L^2_{\Lambda}} \left\{ \left( \frac{\tau^3}{3} \dot{\xi} - \tau^2 \xi\right) \dot{B}^2 + \left( \xi + \tau \dot{\xi} - \frac{\tau^2}{2} \ddot{\xi} - \frac{\tau^3}{3} \dddot{\xi}\right) B^2\right\}\right] \\
& + \epsilon \kappa L_{P} \int \rd \tau \left[ \dddot{\xi} \left( \frac{A^2B^2}{2} - \frac{\epsilon}{L^2_{\Lambda}} \tau^2 B^2\right) - \ddddot{\xi} \frac{\epsilon \tau^3}{6L^2_{\Lambda}} B^2\right]
\end{aligned}
\end{align}
where the first term is a total derivative while the second term vanishes since $\dddot{\xi} =0$. %Computing the symplectic potential, one obtain
%The symplectic potential reads
%\begin{align}
%\Theta & = \epsilon c L_{P} \left[ 2\xi \left( A^2 \dot{B}^2 + 2 A \dot{A} B \dot{B}\right) - \dot{\xi} \left( A \dot{A} B^2 + B \dot{B} A^2\right) \right. \\
%& \left. \qquad \qquad \qquad \qquad \qquad \qquad - \frac{2\epsilon}{L^2_{\Lambda}} \left\{  \left( \frac{\tau^3}{3} \dot{\xi} - \tau^2 \xi\right)\dot{B}^2 +   \left( \tau \xi - \frac{\tau^3}{6} \ddot{\xi}\right)B \dot{B}\right\}\right]
%\end{align}
Then following the Noether theorem, it is straightforward to compute the Noether charge generating the above symmetry. It can be compactly written as
\begin{align}
\begin{aligned}
Q^{\epsilon} %& = F_{\epsilon} - \Theta_{\epsilon} \\
& =  \epsilon \kappa L_{P} \left\{\dot{\xi} \left( A \dot{A} B^2 + B \dot{B} A^2\right) - \ddot{\xi} \frac{A^2B^2}{2} - \xi \left( A^2 \dot{B}^2 + 2 A \dot{A} B \dot{B}\right)  \right\} \\
& \;\; + \frac{\kappa L_P}{L^2_{\Lambda}} \left\{  \left( \frac{\tau^3}{3} \dot{\xi} - \tau^2 \xi\right) \dot{B}^2 + \left( \tau \xi  - \frac{\tau^3}{6}  \ddot{\xi} \right) 2B \dot{B}   - \left( \tau \dot{\xi} - \frac{\tau^2}{2} \ddot{\xi} \right) B^2\right\}
%& = Q^1_{\epsilon} + Q^2_{\epsilon} 
\end{aligned}
\end{align}
The first line corresponds to the Schwarzschild geometry while the second line encodes the corrections induced by the presence of the cosmological constant. Its time evolution can be written as
\begin{align}
\begin{aligned}
\label{consq}
\dot{Q}^{\epsilon} & = \epsilon \kappa L_{P} \left\{ \dot{\xi} B \E_B - 2\xi \left[ \dot{B} \E_{B} +  \dot{A} \E_A \right] - \left[ \frac{A^2B^2}{2}  - \frac{\epsilon}{L^2_{\Lambda}} \left(  \frac{\tau^2}{2} B^2 - \frac{\tau^3}{3} B \dot{B}\right) \right]   \dddot{\xi} \right\}\\
& \qquad \qquad +  \frac{\kappa L_P}{L^2_{\Lambda}} \left\{ \frac{\tau^3}{3} \dot{\xi} - \tau^2 \xi + \tau \xi - \frac{\tau^3}{6} \ddot{\xi}\right\} \frac{2 \dot{B}}{A B}\E_A   \simeq 0
\end{aligned}
\end{align}
where $\simeq$ refers to the on-shell value. Indeed, $\E_B = \E_A \simeq 0$ are the e.o.m for the $A$-field and $B$-field and one has $\dddot{\xi} =0$. The above evolution equation has to be understood as an off shell identity which is the first consequence of the Noether theorem. Imposing the equation of motion then leads to on shell conservation of the quantity $Q_{\epsilon}$ which stands as first integral of the system, i.e a conserved charge. Using the allowed form of the time-dependent function $\xi(\tau)$ given by (\ref{ep}), the three charges generating this $\SL(2,\mathbb{R})$ symmetry are given by (\ref{ch1} - \ref{ch3}).
%\begin{align}
%\label{ch1}
%Q^{\epsilon}_{+} & = -  \epsilon \kappa L_{P} \left\{  A^2 \dot{B}^2 + 2 A \dot{A} B \dot{B} \right\} + \frac{\kappa L_P}{L^2_{\Lambda}} \left\{ 2\tau B \dot{B} - \tau^2 \dot{B}^2\right\}\\
%\label{ch2}
%Q^{\epsilon}_{0} \; & = \epsilon \kappa L_{P} \left\{  A \dot{A} B^2 + B \dot{B} A^2 - \tau  \left( A^2 \dot{B}^2 + 2 A \dot{A} B \dot{B}\right)\right\}  + \frac{\kappa L_P}{L^2_{\Lambda}} \left\{ 2 \tau^2 B \dot{B} - \tau B^2 - \frac{2}{3} \tau^3 \dot{B}^2 \right\}\\
%Q^{\epsilon}_{-} & = \epsilon \kappa L_{P} \left\{2\tau \left( A \dot{A} B^2 + B \dot{B} A^2\right) -  A^2B^2 - \tau^2 \left( A^2 \dot{B}^2 + 2 A \dot{A} B \dot{B}\right)  \right\} \nn \\
%\label{ch3}
%& \qquad \qquad \qquad \qquad \qquad \qquad \qquad \qquad \qquad \qquad \qquad + \frac{\kappa L_{P}}{L^2_{\Lambda}} \left\{  \frac{4}{3} \tau^3 B \dot{B} - \tau^2 B^2 - \frac{\tau^4}{3} \dot{B}^2\right\}
%\end{align}
They are not all independent since they are related by a Casimir condition. In the end, only two of them are needed to solve for the dynamics and therefore to 
label the equilibrium states of the black hole.

%%%%%%%%%%%
Now, we consider a second set of symmetry which corresponds to the translation sector. Consider the transformation
\begin{align}
\label{sym111}
\tau \rightarrow \tilde{\tau} & = \tau \\
\label{sym121}
A \rightarrow \tilde{A}^2(\tilde{\tau}) & =  A^2 (\tau)  + 2h \frac{ \dot{B} }{B} - \dot{h} \\
\label{sym131}
B \rightarrow \tilde{B}(\tilde{\tau}) & =  B (\tau)
\end{align}
Under this transformation, the action becomes
\begin{align}
\label{varTT}
\Delta S % = \frac{L_0L^2_s}{\epsilon L^2_P} \int \rd \tau \left[ \frac{\epsilon}{L^2_s} - \frac{\epsilon}{L^2_{\Lambda}} B^2 - \dot{B}^2 \left( A^2 + 2 h \frac{\dot{B}}{B} - \dot{h} \right) - B \dot{B} \left( 2 A \dot{A} + 2 \dot{h} \frac{\dot{B}}{B} + 2h \frac{\ddot{B}}{B} - 2 h \frac{\dot{B}^2}{B^2} - \ddot{h}\right)\right] \\
%& = S + \frac{L_0L^2_s}{\epsilon L^2_P} \int \rd \tau \left[ - \dot{h} \dot{B}^2 - 2 h \dot{B} \ddot{B} + \ddot{h} B \dot{B}\right] \\
& =  \epsilon \kappa L_{P} \int \rd \tau \left[ \frac{\rd}{\rd \tau} \left( \ddot{h} B^2 - h \dot{B}^2 \right) - \dddot{h} \frac{B^2}{2} \right]
\end{align}
which shows that the action enjoys indeed a Noether symmetry when $\dddot{h} =0$, namely when $h(\tau)$ is quadratic in time, i.e
\be
h(\tau) = \delta +\gamma \tau+ \beta \tau^{2}
\ee
where $(\delta, \gamma, \beta)$ are real constants.
This second symmetry was first found in \cite{Geiller:2020xze} for the Schwarzschild interior model. The above result shows that it is still a symmetry of the Schwarzschild-(A)dS system, both for the exterior and interior regions. The computation of the Noether charges is straightforward. Considering an infinitesimal variation where $h = \xi(\tau)$,  the charges can be compactly written as
\be
T^{\epsilon} = \epsilon \kappa L_{P} \left\{ \xi \dot{B}^2 - \dot{\xi} B \dot{B} + \ddot{\xi} \frac{B^2}{2}\right\}
\ee
Its time evolution vanishes on shell, i.e.
\be
\label{consT}
\dot{T}^{\epsilon} = \epsilon \kappa L_P \left\{ \dddot{\xi} \frac{B^2}{2} - \left( 2\xi \dot{B} - \dot{\xi} B\right) \frac{\E_A}{AB}\right\}  \simeq 0
\ee
where again, $\E_A \simeq 0$ and $\dddot{\xi}=0$. This confirms that $T_{\epsilon}$ provides a second family of evolving constants of motion. The three charges are then easily computed and given by (\ref{cht1} - \ref{cht3}).

\section{M\"{o}bius covariance of the Schwarzschild-(A)dS solution}
%%%

\label{appC}

%Now we would like to translate the Mo\"ebius symmetry of the $\Lambda=0$ theory to a symmetry of the theory at an arbitrary non-vanishing value of the cosmological constant $\Lambda \ne 0$ by using the translation map described above. More precisely, starting with the theory at a non-vanishing value, we would translate it back to the $\Lambda=0$ case, then perform an arbitrary Mo\"ebius symmetry transformation, and finally translate it back to its original value of the cosmological constant. At the end of the day, this means that arbitrary Mo\"ebius  transformations conjugated by the translation map are actual symmetries of the 
Consider the Schwarzschild-(A)dS solution at some value $\Lambda \ne 0$. We first perform a translation $\cT_{h}$, where $h$ is carefully chosen so that $h \triangleright\Lambda=0$. To that end, one can choose the translation functional parameter\footnote{We could also have chosen the function $h_{\Lambda}(\tau)=\tfrac\Lambda3\tau^{3}$, which does not depend on the singularity proper time $\tau_{0}$, or any more general function. Here we made the choice which simplifies as much as possible} as
\be
h_{\Lambda,\tau_{0}}(\tau)\equiv \f\Lambda3(\tau-\tau_{0})^{3}
\,,\qquad
\dddot{h}(\tau)=2\Lambda
\,,\qquad
h(\tau_{0})=\dot{h}(\tau_{0})=0\,.
\ee
Using the notation introduced above, the transformation law of the corresponding translation reads:
\beq
\cT_{h_{\Lambda,\tau_{0}}}\Big{[}A[\tau_0, \tau_1, \mathcal{C}_1, \mathcal{C}_2, \Lambda]\Big{]}^{2}&=&
A[h_{\Lambda,\tau_{0}}\triangleright(\tau_0, \tau_1, \mathcal{C}_1, \mathcal{C}_2, \Lambda)]^{2}
=A[\tau_0, \tau_1, \mathcal{C}_1, \mathcal{C}_2, 0]^{2}
\,,
\\
\cT_{h_{\Lambda,\tau_{0}}}\Big{[}B[\tau_0, \tau_1, \mathcal{C}_1, \mathcal{C}_2, \Lambda]\Big{]}&=&
B[h_{\Lambda,\tau_{0}}\triangleright(\tau_0, \tau_1, \mathcal{C}_1, \mathcal{C}_2, \Lambda)]
=
B[\tau_0, \tau_1, \mathcal{C}_1, \mathcal{C}_2, 0]
\,,\nn
\eeq
with the translation by $h_{\Lambda,\tau_{0}}$ acting simply  on the trajectory's parameter multiplet $(\tau_0, \tau_1, \mathcal{C}_1, \mathcal{C}_2, \Lambda)$ as a simple shift of the cosmological constant:
\be
h_{\Lambda,\tau_{0}}\triangleright(\tau_0, \tau_1, \mathcal{C}_1, \mathcal{C}_2, \Lambda)
=
(\tau_0, \tau_1, \mathcal{C}_1, \mathcal{C}_2, 0)
\,.
\ee
Now we apply a M\"{o}bius transformation by an arbitrary function $f$:
\beq
A[\tau_0, \tau_1, \mathcal{C}_1, \mathcal{C}_2, 0](\tau)
&\underset{\cD_{f}}\longmapsto&
A[f\triangleright(\tau_0, \tau_1, \mathcal{C}_1, \mathcal{C}_2, 0)](f(\tau))=A[\tau_0, \tau_1, \mathcal{C}_1, \mathcal{C}_2, 0](\tau)\,,\qquad
\\
B[\tau_0, \tau_1, \mathcal{C}_1, \mathcal{C}_2, 0](\tau)
&\underset{\cD_{f}}\longmapsto&
B[f\triangleright(\tau_0, \tau_1, \mathcal{C}_1, \mathcal{C}_2, 0)](f(\tau))=\dot{f}(\tau)^{\f12}\,B[\tau_0, \tau_1, \mathcal{C}_1, \mathcal{C}_2, 0](\tau)\,,\nn
\eeq
where the mapping by $f$ leaves the cosmological constant at 0 and acts on the trajectory parameters according to \eqref{eq:Dfparammap}. Then we simply have to perform another translation to switch the cosmological constant back on, by $(h_{\Lambda,f(\tau_{0})})^{-1}(f(\tau))=- h_{\Lambda,f(\tau_{0})}(f(\tau))$ in the time $\ttau=f(\tau)$ (and not in the time $\tau$).
The sequence of three transformations maps the initial trajectory back to itself and thus define a $\SL(2,\R)$ symmetry of the theory, for arbitrary fixed non-vanishing value of the cosmological constant $\Lambda$. To make these transformations more concrete, one can write the resulting composed mapping explicitly:
\be
\label{compo}
\cD_{f}^{\Lambda}\equiv
\cT_{\phi_{\Lambda,f(\tau_{0})}}^{-1}\circ\cD_{f}\circ\cT_{\phi_{\Lambda,\tau_{0}}}
=
\cT_{-\phi_{\Lambda,f(\tau_{0})}}\circ\cD_{f}\circ\cT_{\phi_{\Lambda,\tau_{0}}}
\,.
\ee
The mapping for the $B$-field does not depend on the cosmological constant:
\be
\cD_{f}^{\Lambda}\Big{[}
B[\tau_0, \tau_1, \mathcal{C}_1, \mathcal{C}_2, \Lambda]
\Big{]}(\tau)=
B[f\triangleright(\tau_0, \tau_1, \mathcal{C}_1, \mathcal{C}_2, \Lambda)](f(\tau))
=\dot{f}(\tau)^{\f12}\,B[\tau_0, \tau_1, \mathcal{C}_1, \mathcal{C}_2, \Lambda](\tau)\,,
\ee
where the mapping of the trajectory parameter multiplet $f\triangleright(\tau_0, \tau_1, \mathcal{C}_1, \mathcal{C}_2, \Lambda)$ is the same as before.
On the other hand,  the mapping of the $A$-field is clearly affected by its non-vanishing value:
\beq
\cD_{f}^{\Lambda}\Big{[}
A[\tau_0, \tau_1, \mathcal{C}_1, \mathcal{C}_2, \Lambda]
\Big{]}^{2}
&=&
A[f\triangleright(\tau_0, \tau_1, \mathcal{C}_1, \mathcal{C}_2, \Lambda)]^{2}\circ\tau
\\
&=&
A[\tau_0, \tau_1, \mathcal{C}_1, \mathcal{C}_2, \Lambda]^{2}
+2h_{\Lambda,\tau_{0}}\f{\dot{B}}B-\dot{h}_{\Lambda,\tau_{0}}
-2h_{\Lambda,f(\tau_{0})}\f{\dot{B}^{f}\circ f}{{B}^{f}\circ f} + \dot{h}_{\Lambda,f(\tau_{0})}
\,,\nn
\eeq
where ${B}^{f}\circ f(\tau)=B^{f}(f(\tau))$ stands for $\cD_{f}^{\Lambda}\Big{[} B[\tau_0, \tau_1, \mathcal{C}_1, \mathcal{C}_2, \Lambda] \Big{]}(\tau)$. To be more explicit, it is convenient to write the ratio ${\dot{B}^{f}\circ f}/{{B}^{f}\circ f}$ in terms of $B$ and the derivatives of the time reparametrization function $f$. This gives :
\beq
\cD_{f}^{\Lambda}\Big{[}
A[\tau_0, \tau_1, \mathcal{C}_1, \mathcal{C}_2, \Lambda]
\Big{]}^{2}
=
A[\tau_0, \tau_1, \mathcal{C}_1, \mathcal{C}_2, \Lambda]^{2}
+2h_{\Lambda,\tau_{0}}\f{\dot{B}}B-\dot{h}_{\Lambda,\tau_{0}}
- \frac{2h_{\Lambda,f(\tau_{0})}}{\dot{f}} \left[ \frac{\dot{B}}{B} + \frac{\ddot{f}}{2\dot{f}}\right] +\dot{h}_{\Lambda,f(\tau_{0})}
\,,\nn
\eeq
This expression coincides with the symmetry transformation (\ref{sym11}) introduced in Section~\ref{sec2}.
%\be
%\f{\dot{B}^{f}\circ f}{{B}^{f}\circ f}
%=\f1{\dot{f}}\bigg{[}
%\f{\dot{B}}{B}+\f12\f{\ddot{f}}{\dot{f}}
%\bigg{]}
%\,.
%\ee
This mapping, for arbitrary M\"obius transformation $f$,  defines non-trivial automorphisms on the set of classical trajectories and thus constitutes a symmetry of the Schwarzschild-(A)dS mechanics for an arbitrary value of the cosmological constant\footnote{A subtlety is that we adapted the translations to the parameter $\tau_{0}$. But one can actually allow arbitrary third order polynomials. It would still constitute a symmetry transformation, mapping classical trajectories onto classical trajectories. It would simply imply more intricate transformations for the trajectory parameters.}.

\end{document}